\documentclass[aps,prd,10pt,         
               preprintnumbers,numbers,sort&compress,nofootinbib,
                            showpacs,
               colorlinks,
               linkcolor=blue,
               citecolor=blue]{revtex4-1}
\usepackage{graphicx,amsmath,amssymb,bm}
\usepackage{psfrag}
\usepackage{feynmp}
\usepackage{hyperref}
\usepackage{enumitem}

\usepackage{microtype}

\newcommand{\exclude}[1]{}

\usepackage{epsfig}
\usepackage{dcolumn}

\def\lo{\langle 0 |}
\def\ro{ | 0 \rangle }

\def\atop{ \frac{ \alpha_{s}}{8 \pi} G_{\mu \nu}
 \tilde{G}^{\mu \nu} }

 \def\scalar{\frac{ \alpha_{s}}{8 \pi} G_{\mu \nu}
 {G}^{\mu \nu} }

 \def\la{\langle }
\def\ra{ \rangle }
 \newcommand{\Lqcd}{\Lambda_{\mathrm{QCD}}}
\newcommand{\beq}{\begin{equation}}
\newcommand{\eeq}{\end{equation}}
\newcommand{\bea}{\begin{eqnarray}}
\newcommand{\eea}{\end{eqnarray}}
 
\begin{document}
\title{ Local $\cal P $   Violation Effects and   Thermalization    in QCD:  Views from  Quantum Field Theory and Holography. }

\author{   Ariel R. Zhitnitsky}
 \affiliation{Department of Physics \& Astronomy, University of British Columbia, Vancouver, B.C. V6T 1Z1, Canada}


\begin{abstract}
We argue   that the local violation of  ${\cal P}$ and ${\cal CP}$ invariance in heavy ion collisions and  the universal thermal aspects observed in   high energy collisions are in fact two sides of the same coin, and both are related to quantum anomalies of QCD.   We argue that the   low energy relations   representing  the quantum anomalies of QCD are saturated by  coherent  low dimensional   vacuum configurations
as observed in Monte Carlo lattice studies.   
The thermal  spectrum and  approximate universality of the temperature with no dependence on energy of colliding particles  in this framework  is due to the fact that the emission   results from the distortion   of these low dimensional   vacuum sheets  rather than from the colliding particles themselves.
The emergence of the    long- range correlations  of ${\cal P}$  odd domains (a feature which is apparently  required  for explanation  of the  asymmetry  observed at RHIC and LHC) is also a result of the same distortion of the QCD vacuum configurations. We   formulate the corresponding physics using  the effective low energy effective Lagrangian.  We also formulate the same physics  in terms of the dual holographic picture  when low-dimensional sheets of topological charge embedded in 4d space, as observed in Monte Carlo simulations, are identified with D2 branes. 
Finally, we argue that  study of these  long range correlations in heavy ion collisions could serve as a perfect test of a proposal that the observed dark energy 
in present epoch  is a result of a tiny deviation of the QCD vacuum energy in expanding universe from  its conventional   value in Minkowski  spacetime.
  
   \end{abstract}
\maketitle

\section{Introduction. Motivation}\label{motivation}
Recently it has become clear that quantum anomalies (chiral and conformal)  play  very important role in the macroscopic dynamics of relativistic fluids. Much of this progress is motivated by very interesting ongoing  experiments 
on local ${\cal{P}}$  and ${\cal{CP}}$ violation in QCD as studied  at RHIC and ALICE at the LHC~\cite{Voloshin:2004vk,Abelev:2009tx,collaboration:2011sma,Selyuzhenkov:2011xq}. It is likely that the observed asymmetry is due to   charge separation effect  \cite{Kharzeev:2004ey,Kharzeev:2007tn}  as a result of the chiral anomaly, see details below. 

Another, naively unrelated, but well established property of QCD, apparently also related to the quantum anomaly,  can be formulated    as follows. Over the years, hadron production studies in a variety of high energy collision experiments have shown a remarkably universal feature, indicating a universal hadronization temperature $T_H\sim (150- 200) ~{\text MeV}$.
  From $e^+e^-$  annihilation to $pp$ and $p\bar{p}$  interactions and further to collisions of heavy nuclei, with energies from a few GeV up to the TeV range, the production pattern always shows striking thermal aspects, connected to an apparently quite universal temperature around $T_H\sim (150- 200) ~{\text MeV}$ \cite{hagedorn}. It has been recently argued that this feature may  also be related to the quantum anomaly,   the conformal anomaly ~\cite{Castorina:2007eb}. The basic idea behind this bold proposal is the observation that the effective Lagrangian 
 which exactly accounts for the effects of the conformal anomaly indicates a presence of  the event horizon ~\cite{Castorina:2007eb}, which eventually leads 
  to the universal thermal Hawking-Unruh radiation, see details below. 
  
Few arguments  that these two naively unrelated phenomena  may   in fact  be originated from the  same non-perturbative QCD dynamics were recently presented  in \cite{Zhitnitsky:2010zx}. However, the arguments  of ref.~\cite{Zhitnitsky:2010zx} were based on an effective description 
in terms of an  acceleration $``\mathbf{a}"$ of a Rindler observer, while microscopical picture in terms of specific  gauge configurations responsible for the effect was completely obscured in formulation  \cite{Zhitnitsky:2010zx}. The goal of the present paper is (at least partially) to  fill this gap, and develop some technical tools which might be  appropriate to address those hard questions.

Before we develop  a suitable  technique to deal with both  these two phenomena, we review each effect separately as it is 
commonly  treated  today. Our next step is 
to construct the low energy effective Lagrangian which accounts for the quantum anomalies.  The quantum anomalies  are normally represented in a form of specific properties  of two point correlation functions. We opted, however to code this information in a form of low energy effective Lagrangian.  It allows us to study the relevant non-perturbative gauge configurations which are responsible for saturation  of the low energy theorems representing the quantum anomalies. The corresponding analysis will be supported by  some recent lattice results as well as holographic interpretation of these lattice results.

  \subsection{Local ${\cal{P}}$  and ${\cal{CP}}$ violation in QCD.  Charge separation effect}\label{P}
 We start with review of the  charge separation effect  \cite{Kharzeev:2004ey,Kharzeev:2007tn} which can be explained in the following simple way. Let us assume that an  effective   $\theta (\vec{x}, t)_{ind}\neq 0$  is induced as a result of some non- equlibrium dynamics as suggested in refs.
 \cite{Kharzeev:1998kz,Kharzeev:1999cz,Buckley:1999mv,Buckley:2000aa}. The  $\theta (\vec{x}, t)_{ind}$ parameter enters the effective lagrangian as follows,  ${\cal L_{\theta}}=-\theta_{ind} q$ where $ q \equiv \frac{g^2}{64\pi^2} \epsilon_{\mu\nu\rho\sigma} G^{a\mu\nu} G^{a\rho\sigma}$ such that local ${\cal{P}}$  and ${\cal{CP}}$
  invariance of QCD is broken on the scales where correlated $\theta (\vec{x}, t)_{ind}\neq 0$  is induced. As a result of this violation, one should expect a number of ${\cal{P}}$  and ${\cal{CP}}$ violating effects taking place in the region where $\theta (\vec{x}, t)_{ind}\neq 0$.  
  \exclude{In particular, one should expect the separation of electric charge along the axis of magnetic field $\vec{B}$ or along the angular momentum $\vec{l}$
  if they are present in the region with $\theta (\vec{x}, t)_{ind}\neq 0$. }
   
  This area of research  became a very active  field in recent years mainly due to very interesting ongoing  experiments~\cite{Voloshin:2004vk,Abelev:2009tx,collaboration:2011sma,Selyuzhenkov:2011xq}.    There is a number of different manifestations of this  local ${\cal{P}}$  and ${\cal{CP}}$  violation, see  \cite{Kharzeev:2007jp,Fukushima:2008xe,Kharzeev:2009fn,Buividovich:2009wi,Abramczyk:2009gb} and many additional references therein. In particular, in the presence of an external magnetic field $\vec{B}$ or in case of the rotating  system  with angular velocity $\vec{\Omega}$ there will be induced electric   current directed along $\vec{B}$ or $\vec{\Omega}$ correspondingly, resulting in   separation
 of charges along those directions as mentioned above. One can interpret the same effects as a generation of induced electric field $\vec{E}$ directed along $\vec{B}$ or $\vec{\Omega}$  resulting in corresponding electric current flowing along $\vec{J}\sim \vec{B}$ or 
 $\vec{J}\sim \vec{\Omega} $   directions. All these phenomena are  obviously   ${\cal{P}}$  and ${\cal{CP}}$
 odd effects. 
Non-dissipating, induced vector current  density    has the form:
  \beq
  \label{J}
  \vec{J}=(\mu_L-\mu_R)\frac{e\vec{B}}{2\pi^2},
  \eeq
  where ${\cal{P}}$ odd effect is explicitly present in this expression as the difference of chemical potentials of the right $\mu_R$ and left $\mu_L$ handed 
  fermions is assumed to be nonzero, $(\mu_L-\mu_R)\neq 0$ in the region where $\theta (\vec{x}, t)_{ind}\neq 0$. The combination $(\mu_L-\mu_R)$ can be thought
  as $\dot{\theta} (t) $ after  a corresponding $U(1)_A$ chiral time-dependent rotation is performed, see also~\cite{Kharzeev:2009fn}
  for a physical interpretation of the relation $(\mu_L-\mu_R)={\dot{\theta }(t)_{ind}}$.
   It is important to emphasize that the 
  region where  $\la\theta (\vec{x}, t)_{ind}\ra\neq 0$  should be much larger in size than the scale of  conventional QCD fluctuations with   correlation length $\sim \Lambda_{QCD}^{-1}$.

  Closely related phenomena have been previously discussed in  the physics of neutrinos \cite{vilenkin}
  and quantum wires \cite{alekseev}.  In QCD context formula (\ref{J})
  has been used in applications to neutron star physics where magnetic field is known to be large, and the corresponding $(\mu_L-\mu_R)\neq 0$ can be  generated   in neutron star environment as a result of  continuos ${\cal{P}}$  violating processes happening in nuclear matter~\cite{Charbonneau:2007db,Charbonneau:2009ax}. It has been also applied to heavy ion collisions where an effective $(\mu_L-\mu_R)\neq 0$ 
  is locally induced. The effect  was estimated using the sphaleron transitions generating the topological charge density in the QCD plasma \cite{Kharzeev:2007jp,Fukushima:2008xe}. The effect was  coined as ``chiral magnetic effect" (CME)~\cite{Kharzeev:2007jp,Fukushima:2008xe}.  Formula (\ref{J}) has been also derived a numerous number of times using varies  techniques such as: effective lagrangian approach  developed in \cite{Son:2004tq}; explicit mode's summation
    \cite{Metlitski:2005pr}; direct lattice computations \cite{Buividovich:2009wi,Abramczyk:2009gb}. 
   In addition, the effect has been studied in holographic models of QCD~\cite{ads/cft,Brits:2010pw}.  
   
    To summarize: on the theoretical side the effect is  well established phenomenon.  Future experiments at RHIC and LHC   hopefully will  teach  us much more on implementation of this effect   to  heavy ion collisions, see  review ~\cite{Kharzeev:2011vv} with large number of references on recent developments. 
    Furthermore, as all measurements   are in fact ${\cal P}$ even observables, there are many background processes  which considerably contribute to the effect
    \cite{Bzdak:2009fc,Liao:2010nv,Bzdak:2010fd,Schlichting:2010qia,Pratt:2010zn}, see some comments on this matter in section  \ref{collisions}. Therefore, it remains to be seen if the CME/charge separation effect  is a main  source of the observed event-by-event fluctuations. In this work we assume this is to be the case, and interpret the observations  at RHIC and the LHC  as a manifestation  of the CME.
    
    One of the crucial questions for the applications of the CME to heavy ion collisions is  a correlation length of the induced $\la\theta (\vec{x}, t)_{ind}\ra\neq 0$: why the ${\cal{P}}$  odd domains are  large, much larger than conventional 
       $\Lambda_{QCD}^{-1}$ scale?  Apparently, a relatively  large correlation length is a required feature for interpretation  of the observed  asymmetry 
       \cite{Voloshin:2004vk,Abelev:2009tx,collaboration:2011sma,Selyuzhenkov:2011xq} in terms of CME
       as the conventional QCD vacuum processes  are too small to explain the observed asymmetry~\cite{Asakawa:2010bu}.  This is  in fact,  the key element to be addressed in this paper: why the  correlation length of $\cal{P}$ odd fluctuations could be large in heavy ion collisions? 
      
  \subsection{Universal hadronization temperature $T_H\sim (150- 200) ~{\text MeV}$}\label{TT}
 The second  part of the story goes as follows. 
  While experimentally universal temperature around $T_H\sim (150- 200) ~{\text MeV}$ \cite{hagedorn} 
   is well established phenomenon,  its origin is difficult to understand as number of incident particles in $e^+e^-$  annihilation as well as in $pp$ and $p\bar{p}$  interactions is not sufficient even to talk about statistical averages. This observation motivated a number of  early attempts \cite{T_early} to interpret the resulting spectrum of particles as the Hawking- Unruh radiation~\cite{Birrell:1982ix} when the event horizon emerges as a result of strong interactions
     with an effective temperature 
  \beq
  \label{T}
  T=\frac{ \mathbf{a}}{2\pi}\sim \Lambda_{QCD},
  \eeq 
 being expressed in terms of   effective deceleration parameter $``\mathbf{a}"$. 
   Different aspects of this idea  were advocated in a number of recent papers, see e.g. \cite{Castorina:2007eb,Kharzeev:2005iz}. 
   The   problem of calculating of the effective parameter  $``\mathbf{a}"$ is obviously very hard problem of strongly interacting QCD.

      One should emphasize that the Planck spectrum in this approach is not  resulted   from the  kinetics when the thermal equilibrium with temperature $T$ given by eq. (\ref{T}) is reached  due to the large number of collisions. Rather, the Planck  spectrum in high energy collisions in this framework  is resulted from the stochastic tunnelling  processes when no information transfer occurs. In such circumstances   the spectrum must be thermal. Such interpretation would naturally explain both puzzles of the phenomena: \\
      a)  the thermal spectrum in $e^+e^-, ~pp$ and $p\bar{p}$ high energy collisions emerges in spite  of the fact that the statistical thermalization could  never be reached in those systems; \\
b) an approximate  universality of the temperature as it is expressed in terms of unique and fundamental scale
 $ \Lambda_{QCD} $   entering formula (\ref{T}) with no dependence on energy of colliding particles in this formula.

  The paper is organized as follows.  Section \ref{effective}
is devoted to the discussions of low energy relations in pure gluodynamics with no quarks.
 The corresponding relations can be conveniently represented in terms of the low energy Lagrangian
 for the background dilaton $\eta (x) $ and  axion $a(x)$ fields combined into a single complex field which describes the dynamics of scalar $\eta\sim \scalar$ and pseudo scalar $a\sim \atop $ components of the underlining  gauge theory. 
    This will allow us to study the long range  effects 
  of gauge configurations on scales which are 
  much larger than conventional 
       $\Lambda_{QCD}^{-1}$ scale. The basic reason for emergence of this long range order  is 
       the $2\pi$ periodic properties of the axion field $a(x)$ which is a crucial feature of QCD. This periodicity may result in formation of the topological solitons 
       characterized by long range order. 
       In section \ref{lattice}
       we discuss  some recent results from lattice simulations and holographic description as they seen from 
       effective Lagrangian description. We shall identify the long range structure observed in lattice simulations with D2 branes in holographic description and with the axion field solitons in the effective Lagrangian approach. 
         In section \ref{applications} we discuss some applications of these ideas. Specifically,  in section \ref{collisions}  we discuss  local $\cal{P}$ violation effects and universality of  hadronization  temperature $T_H$ in high energy collisions,  while  in section \ref{cosmology} we discuss applications to cosmology.
  
   \section{Effective Lagrangian in gluodyamics}\label{effective}
   In next subsection \ref{theorems} we review the  low energy relations in gluodynamics. 
   In subsection \ref{lagrangian} we construct the low energy effective  lagrangian which is generating functional for these low energy correlation functions.
 \subsection{Low energy relations }\label{theorems}

In what follows we consider two types of low energy correlation functions  (scalar and pseudoscalar)   in gluodynamics. 
For the scalar channel case, it was shown long ago  
in ref. \cite{NSVZ} that these correlation functions 
are fixed by renormalizability and conformal
anomaly in gluodynamics, 
\beq
\label{3}
\lim_{q \rightarrow 0 } \; i 
\int dx \, e^{iqx} \lo T 
\{ \frac{\beta(\alpha_s)}{4 \alpha_s} G^2 (x)  \,  
 \frac{\beta(\alpha_s)}{4 \alpha_s} G^2 (0) \} \ro  = 
-4  \la \frac{
\beta(\alpha_s)}{ 4 \alpha_s} G^2 \ra \; ,
\eeq
where the one-loop $ \beta $-function, $ \beta(\alpha_s) = 
- b \alpha_{s}^2 /(2 \pi) $ 
 with
 $ b = (11/3) N_c $ and $ N_c $  stands for the number of colours.
  Arbitrary n-point functions of the trace of the energy-momentum
tensor 
\beq
4E_{\text vac}\equiv \la \sigma \ra = \la \frac{\beta(\alpha_s)}{4 \alpha_s} \,  G^2
\ra = \la \frac{- b \alpha_s }{8 \pi} \, G^2 \ra  + O(\alpha_{s}^2 ) 
\eeq 
can be obtained by further 
differentiating relation (\ref{3}) :
\beq
\label{5a}
i^n \int dx_1 \ldots d x_n \lo T \{ \sigma (x_1) \ldots \sigma
(x_n) \, \sigma (0) \} \ro = ( - 4)^n  \, \la \sigma \ra \; , 
\eeq
where, as in Eq.(\ref{3}), a limiting procedure 
of the vanishing momentum transfer $ q_{\mu} $ is implied.
  
 Let us now address zero momentum 
correlation functions of the topological density operator
in gluodynamics, the so-called topological susceptibility $\chi$. It can be written as
\bea
\label{5}
\chi &\equiv& -\frac{\partial^2 E_{\text vac}}{\partial \theta^2}=\lim_{ q \rightarrow 0} \; 
i \int dx \, e^{iqx} \lo T \left\{ \frac{\alpha_s}{8 \pi} 
G \tilde{G} (x)  \, 
\frac{\alpha_s}{8 \pi} G \tilde{G} (0) \right\}  \ro    \nonumber  \\
&=& \xi^2  \, \la \frac{
\beta(\alpha_s)}{ 4 \alpha_s} G^2 \ra =\frac{1}{N_c^2} E_{\text vac}(\theta)\; ,
\eea
where $ \xi $ stands for a generally unknown numerical coefficient
(note that its $N_c $ dependence is expected to be 
$ \xi \sim N_{c}^{-1} $, in order to match Witten-Veneziano
\cite{ven,vendiv,witten} resolution of the $U_A (1)$ problem, see also \cite{Rosenzweig:1979ay,Nath:1979ik,Kawarabayashi:1980dp}). 
It has been argued in 
 \cite{Halperin:1997bs,Halperin:1998rc} that the corresponding consistent effective Lagrangian can be constructed for any rational $\xi=q/p$.
 However, in what follows, without loosing any generality, we consider the simplest and most appealing case when $\xi=\frac{1}{2N_c}$ which corresponds to 
  $E_{\text vac}(\theta)\sim \cos(\theta/N_c)$ behaviour for the vacuum energy.  Such a behaviour is realized in all supersymmetric cases. 
  The same  $ \cos(\theta/N_c)$ structure emerges even when supersymmetry is slightly broken, see details and references in  \cite{Halperin:1997bs,Halperin:1998rc}. Furthermore, such a behaviour is also supported by  the Veneziano construction \cite{ven}
  when the Veneziano ghost saturates all relevant $2n-$ point correlation functions generalizing (\ref{5})
   \bea
\label{6}
  \frac{\partial^{(2n-1)}}{\partial \theta^{(2n-1)}}
  \left< q(0)\right>\sim 
  \int dx_1 \ldots d x_{2n-1} \lo T \{ q (x_1) \ldots q
(x_{2n-1}) \, q(0) \} \ro
  \sim\frac{E_{\text vac}}{N_c^{2n}}.
\eea
Most importantly, the same behaviour can be   demonstrated to emerge \cite{Thomas:2011ee} in four dimensional ``deformed QCD" formulated in \cite{Yaffe:2008} where all computations can be explicitly performed as the model is in a weak coupling regime. In the ``deformed QCD" 
the behaviour (\ref{6}) is direct manifestation of topological structure of the ground state when 
the non-trivial topological sectors of the theory and transitions between them are described in terms of the weakly coupled monopoles. 
\exclude{The monopoles  live in Euclidean space and describe the physical tunnelling processes between different topological sectors $|n\ra$ and $| n+1 \ra$. The corresponding tunnelling processes saturate the correlation functions (\ref{6})  and the vacuum energy $E_{\text vac}$ precisely in the way to satisfy (\ref{6}). }The basic reason why eq. (\ref{6}) holds is that all the correlation functions as well as the vacuum energy $E_{\text vac}$ are saturated by the same vacuum fluctuations which have non-dispersive nature, contribute to  (\ref{5}), (\ref{6}) with ``wrong sign", and can not be associated with any physical propagating degrees of freedom,  see \cite{Thomas:2011ee} for the technical details.

We note finally  that the correlation 
functions (\ref{5}), (\ref{6}) are  defined via the path integral, i.e. with 
Wick's  T-product in contrast with conventional Dyson's T-product when only physical asymptotic states 
contributes to the corresponding correlation functions, see Appendix   of ref. \cite{Zhitnitsky:2011tr} for discussions on some subtleties in the energy definition.

\subsection{Effective Lagrangian  }\label{lagrangian}

The purpose of this section is to review the construction of the 
low energy effective Lagrangian for gluodynamics~\cite{Halperin:1997bs,Halperin:1998rc}.
The corresponding  Effective Lagrangian 
 contains all information provided by the low energy 
relations in the scalar (\ref{3}) and pseudoscalar (\ref{5}) channels
including all multi-point correlation functions of operators 
$ G^2 $ and $ G \tilde{G} $, which can be obtained by differentiating
the two-point functions (\ref{3}) and (\ref{5}) with respect to $1/g^2$ and $\theta$, 
see e.g. Eq.(\ref{5a}), (\ref{6}).

Before proceeding with the presentation, we would like to pause
for a comment on the meaning of this effective Lagrangian. As 
there exist no Goldstone bosons in pure gluodynamics, no Wilsonian
effective Lagrangian, which would correspond to integrating out 
heavy modes, can be constructed for gluodynamics. Instead, one 
speaks in this case of an effective Lagrangian as a generating 
functional for vertex functions of the composite fields $ G^2 $ and 
$ G \tilde{G} $. Moreover, only the potential part of this 
Lagrangian can be computed  as it corresponds to zero momentum n-point
functions of $ G^2 , G \tilde{G} $, fixed by the low energy theorems.
  Thus, such an effective Lagrangian is not very useful
for calculating the S-matrix elements, but is perfectly suitable for addressing 
the vacuum properties of the theory, including the coordinate dependent vacuum configurations 
if they are treated as the background fields, see next section.

The task of constructing an effective 
Lagrangian can be considerably simplified as suggested in ~\cite{Halperin:1997bs,Halperin:1998rc} 
by going over to linear combinations of original 
operators
\exclude{\footnote{
In this section we change the normalization of the gluon field 
in comparison to that used in Sect.\ref{theorems} by the rescaling $ A_{\mu} 
\rightarrow (1/g) A_{\mu} $.}   }
which enter relations (\ref{3}), (\ref{5}) :
\beq
\label{9}
H  =  \left(-\frac{b}{64 \pi^2}   
G^2  + i \, \frac{N_c}{16 \pi^2} \ 
G \tilde{G} \right) \; , \; \bar{H}   
= \left(-\frac{b}{64 \pi^2}   
G^2  - i \, \frac{N_c}{16 \pi^2} \ 
G \tilde{G} \right) \;  .
\eeq  
In terms of these combinations, the low energy relations 
for renormalized zero momentum Green function, Eqs. (\ref{3}) 
and (\ref{5}), take particularly simple forms (for an arbitrary
value of the vacuum angle $ \theta $):
\bea
\label{10}
\lim_{q 
\rightarrow 0 } \, i \int dx 
e^{iqx} \lo T \{ H(x) \; H(0) \} \ro &=& - 4 \la H \ra \; , 
\nonumber \\
\lim_{ q \rightarrow 
0} \,  i \int dx 
e^{iqx} \lo T \{ \bar{H}(x) \; \bar{H}(0) \} \ro &=& - 4 
\la \bar{H} \ra \; , \\ 
\lim_{ q \rightarrow 
0} \,  i \int dx 
e^{iqx} \lo T \{ \bar{H}(x) \; H(0) \} \ro &=&  0 \; . \nonumber 
\eea
It is easy to check  that the decoupling of the fields $ H $ 
and $ \bar{H} $ 
holds for arbitrary n-point functions 
of $ H $, $ \bar{H} $. This circumstance makes it particularly 
convenient to work with fields (\ref{9}). 

We now formulate  the  effective low energy Lagrangian 
reproducing at the 
tree level all  low energy relations 
for the composite fields $ H , \bar{H} $ discussed in previous subsection~\ref{theorems},
\bea
\label{pot}
e^{- i V F(h,\bar{h}) } &=& \sum_{n = - \infty}^{
 + \infty}   \exp \left\{ - \frac{i V}{4}
\left( h \, \ln \, \frac{h}{2 e |E_{\text vac}| } + 
\bar{h} \, \ln \, \frac{ \bar{h}}{
 {2 e |E_{\text vac}| }} \right) \right. \nonumber \\ 
&+& \left. i \pi V \left(  
\frac{ \theta + 2 \pi n}{ 2 \pi N_c} \right) \frac{h - \bar{h}}{
2 i} \right\} \;  ,
\eea
where $V$ is the 4-volume of the system, and the effective zero momentum fields $h, \bar{h}$ satisfy the equations
\beq
\int dx ~h=\la \int dx ~H \ra, ~~~~ \int dx ~\bar{h}=\la \int dx ~\bar{H} \ra .
\eeq
 For the case of single ``dilaton" field $ \sigma = - b \alpha_s/
(8 \pi) \, G^2 $, a similar problem of constructing an effective 
Lagrangian was solved long ago  ~\cite{MS}, see also some applications of these ideas in given context in refs~\cite{Kharzeev:2002rp,Kharzeev:2004ct,Kharzeev:2008br}.    To reproduce the dilaton potential constructed in  ~\cite{MS} 
one should parametrize   $h=\bar{h}=2  |E_{\text vac}|\exp(\eta)$ in formula (\ref{pot}) to arrive to
\beq
\label{MS}
F(\eta)= -|E_{\text vac}| e^{\eta} (1-\eta), ~~~ 4E_{\text vac}\equiv \la \sigma \ra =   \la \frac{- b \alpha_s }{8 \pi} \, G^2 \ra .
\eeq
The  minimum of the potential is positioned at $\eta_{min}=0$ which   determines the ground state of the system
to be $F_{min} (\eta=0)=-|E_{\text vac}|$.

The generalization of this construction was suggested in ~\cite{Halperin:1997bs,Halperin:1998rc}.  It is expressed by formula (\ref{pot}) and contains two new elements. First, it has an additional axion\footnote{We use term ``axion" which is 
a jargon here.  There is no real new dynamical degree of freedom such as axion, see recent reviews~\cite{axion} about   physical axion as real degree of freedom. However, the vacuum expectation value of $a$ field in $\theta$ vacua  is directly related to $\theta$ as follows $\la a\ra=\theta/N_c$, see eq. ~(\ref{min}) which justifies  our terminology as the dynamical $\theta(x)$ is, by definition, the axion.}  field $``a"$ along with the dilaton field $\eta$ 
defined as follows:
\beq
\label{h}
h = 2 |E_{\text vac}| \, e^{ \eta + i a} \; \; , \; \; 
\bar{h} = 2 |E_{\text vac}| \, e^{ \eta - i a } \; .
\eeq
This new $``a"$  field describes the dynamics of the pseudoscalar component $  G \tilde{G}  $ represented by  low energy relations 
(\ref{5}),(\ref{6}).
The second new element is that the structure of 
the effective potential $ F(h,\bar{h}) $ is such that it contains
along  with conventional  ``dynamical" part represented by the first term in the exponent in eq. (\ref{pot}) 
also the  ``topological" part represented by  the second term in the exponent in eq. (\ref{pot}).  
The conventional  ``dynamical" part  is merely  a kinematical
reformulation of the content of the low energy relations (\ref{10}).
The ``topological" term on other hand does not modify the local relations (\ref{10}), but rather, 
 guarantees the $2\pi$ periodicity $\theta\rightarrow\theta+2\pi n$
 of the energy of the ground state of the system.  Indeed, the minimization of the effective potential (\ref{pot})
 leads to the following result for the ground state ~\cite{Halperin:1997bs,Halperin:1998rc}:
 \bea
\label{min}
F_{min} (\theta)&=&  - \lim_{V \rightarrow 
\infty} \; \frac{1}{V} \ln \left\{ 
  \sum_{n=0}^{N_c-1} \exp \left[ 
V|E_{\text vac}| \cos \left( \frac{ \theta + 2 \pi \, n }{N_c} \,\right) \right]   \right\},\\
\nonumber  ~~\eta_{min}&=&0, ~~a_{min}= \frac{ \theta + 2 \pi \, l }{N_c}.
\eea
The vacuum energy $F_{min} (\theta)$ is obviously $2\pi$ periodic function in spite of the fact that $\theta$ parameter enters the expression for the energy as $\theta/N_c$ in accordance with Witten-Veneziano resolution of the $U_A(1)$ problem represented by formula (\ref{5}).  This important property  is achieved due to the ``topological" part in effective potential (\ref{pot}). This term ensures that the effective potential 
has not one, but rather $ N_c $ physically different local extrema.
  The physics is perfectly 
periodic\footnote{Such a pattern is known to emerge in many four dimensional supersymmetric models, and also gluodynamics in the limit $N_c=\infty$. It has been further argued ~\cite{Halperin:1997bs,Halperin:1998rc} that the same  pattern also emerges in four dimensional gluodynamics at any finite $N_c$.    The same pattern emerges in holographic description of QCD ~\cite{wittenflux} at $N_c=\infty$ as well. Finally, this pattern is realized in  weakly coupled ``deformed QCD" model where all computations are under complete theoretical control \cite{Thomas:2011ee}.
 } in $ \theta $ with period
$ 2 \pi $, as the minima 
interchanging under the shift 
$ \theta \rightarrow \theta + 2 \pi $.
  At the same time, we observe  
level crossing with a two-fold degeneracy at 
 $ \theta = 
\pi \; (mod \, 
2 \pi ) $.  
The construction of the potential similar to (\ref{MS}) when the axion field is taken  into account leads to the following expression 
\bea
\label{potential}
F(\eta(x), a(x))= -  |E_{\text vac}| \left[e^{\eta} (1-\eta)\cos a +e^{\eta}a\sin a \right], 
\eea 
where we keep only the lowest branch in   expression  (\ref{pot}) to simplify notations. 
This formula  reduces, of course, to  previous expression  (\ref{MS}) in the limit $a\rightarrow 0$ when the axion field is neglected. One should remark here that the expression for $F(\eta, a)$ is  not  literally periodic $a\rightarrow a+2\pi$
because $ F(\eta(x), a(x)) $ in eq. (\ref{potential}) corresponds to the lowest branch with $l=0$. 
However, the $2\pi$ periodicity is restored when summation over all branches in partition function is implemented, similar to our procedure leading to eqs.(\ref{pot}), (\ref{min}).

The corresponding  discrete set of degenerate vacuum states as a result of the  $2\pi $ periodicity of the effective potential (\ref{pot}) for the axion field  is a signal that the domain  wall configurations 
interpolating between these   states are present in the system.  However, the corresponding configurations are not conventional domain walls similar to the well known ferromagnetic domain walls in condensed matter physics which interpolate between physically {\it distinct} vacuum states. In contrast, in present case a corresponding  configuration interpolates between 
      topologically different but physically equivalent  winding states $| n\ra$, which are connected  to each other by large gauge transformation operator, see next section with elaboration on this issue. Axion field acts as a probe to signal that the degeneracy is present in the system and the presence of the  domain walls is expected in the system. 
      The corresponding domain wall configurations  in Euclidean space will be interpreted in next section as configurations describing the tunnelling processes in Minkowski space, similar to  Euclidean instantons. This interpretation should  be  contrasted with 
     conventional  interpretation of  static domain walls defined    in    Minkowski space. 
    Most important  lesson from these discussions    is as follows.  The effective potential (\ref{pot}) which is generating functional 
    for  the low energy relations hints on the presence of  the unusual topological configurations due to its generic $2\pi$ periodicity.
    The physical meaning of these configurations will be elaborated  later in the text.

   \section{Insights from Lattice Simulations and from Holographic Picture of QCD }{\label{lattice}
 In this section we want to get some insights on crucial vacuum configurations from the   lattice results. The Monte Carlo simulations are normally performed in Euclidean space. Therefore, we reformulate the low energy relations discussed in previous section  \ref{effective}  to Euclidean space time in order to make comparison with lattice results. 
 \subsection{Topological susceptibility}\label{qft}
 The scalar correlation function in Euclidean space takes the form and it is negative
 \beq
\label{Euclidean-1}
 \int dx \,  \lo T \left\{ \frac{b\alpha_s}{8 \pi} 
G^2 (x)  \, 
\frac{b\alpha_s}{8 \pi} G^2 (0) \right\}  \ro  = -4 \lo\frac{b\alpha_s}{8 \pi} G^2  \ro ~~ <0, 
\eeq
while the topological susceptibility in the Euclidean space is positive
\bea
\label{Euclidean-2}
\chi_{Eucl}  =  
 \int dx \,  \lo T \left\{ \frac{\alpha_s}{8 \pi} 
G \tilde{G} (x)  \, 
\frac{\alpha_s}{8 \pi} G \tilde{G} (0) \right\}  \ro    
  =\frac{1}{N_c^2} |E_{\text vac}(\theta)| ~~ >0.
\eea
The difference in signs\footnote{ A warning signal with the signs: the physical degrees of freedom in  Euclidean space (where the lattice computations are performed)  contribute  to topological susceptibility $\chi_{QCD}$ with the negative sign, while the contact term (the Veneziano ghost) contributes with the positive sign, in contrast with our discussions in Minkowski space, see eqs. (\ref{3}), (\ref{5}).} between these two correlation functions can be seen in Minkowski space as well, see eq. (\ref{3}) versus (\ref{5}). The crucial observation here is as follows: any physical state contributes to $\chi_{Eucl}$  with negative sign 
\bea
\label{dispersion}
\chi_{dispersive} \sim  \lim_{k\rightarrow 0} \sum_n  \frac{\la 0|q  |n\ra \la n | q |0\ra }{-k^2-m_n^2} <0,
\eea
in drastic contrast with low energy relation (\ref{Euclidean-2}). It poses no problem for the correlation function (\ref{Euclidean-1}) when the physical dilaton saturates the negative sign in eq.(\ref{Euclidean-1}).
At the same time the positive physical mass $m_{\eta'}^2 > 0$ for the $\eta'$ meson requires 
the  positive sign for the topological susceptibility  (\ref{Euclidean-2}), see the original reference~\cite{vendiv} for a thorough discussion.
Therefore, there must be a contact  contribution to $\chi$, which is not related to any propagating   physical degrees of freedom,  and it must have a ``wrong sign" (in comparison with (\ref{dispersion}) representing the conventional dispersive contribution) to saturate  the positive sign for topological susceptibility (\ref{Euclidean-2}). 
In different words, it must be a non-dispersive contribution to $\chi$ which is not associated with any asymptotical physical states in conventional dispersion relations.  In the framework  \cite{witten} 
the  contact term with  ``wrong sign" has been postulated, while in refs.\cite{ven,vendiv} the Veneziano ghost had been  introduced to saturate the required 
property (\ref{Euclidean-2}). 

The simplest way to convince oneself in necessity for a non-dispersive contribution to $\chi$ with a ``wrong sign" 
 is to  compute 
the topological susceptibility $\chi_{QCD}$ in QCD rather than in gluodynamics.  
The topological susceptibility $\chi_{QCD} (m_q= 0)=0 $ must vanish in the chiral limit as a consequence of the Ward Identities (WI).  It is very instructive to see  how it happens. 
If one models the contact contribution to $\chi$ using the  Veneziano ghost, the topological susceptibility in Euclidean  space can be  represented as 
follows, see \cite{Zhitnitsky:2010zx, Zhitnitsky:2011aa} and references therein: 
 \bea
\label{top}
\chi_{QCD}\equiv \int \! d^4x \lo T\{q(x), q(0)\}\ro_{QCD}= \frac{f_{\eta'}^2 m_{\eta'}^2}{4} \cdot \int d^4x\left[ \delta^4 (x)- m_{\eta'}^2 D^c (m_{\eta'}x)\right]
\eea
where $D^c (m_{\eta'}x)$ is the Green's function of a free massive particle with standard normalization $\int d^4x m_{\eta'}^2 D^c (m_{\eta'}x)=1$. 
The term proportional $ -D^c (m_{\eta'}x)$ with negative sign  in eq. (\ref{top}) is  resulted from the lightest  physical $\eta'$ state  of   mass $m_{\eta'}$ and it has a negative sign in accordance with (\ref{dispersion}).
At the same time  the $\delta^4(x)$ represents the ghost contribution with   ``wrong"  sign 
which can not be associated with any physical states. The ghost's contribution can be also thought as the Witten's contact term~\cite{witten} not related to any propagating degrees of freedom. The topological susceptibility $\chi_{QCD} (m_q= 0)=0 $ vanishes in the chiral limit  as a result of exact cancellation between two terms entering (\ref{top}) in complete accordance with   WI.    The WI can not be satisfied if the contact term is not present in the system. 
When $m_q\neq 0$ the cancellation is not complete and $\chi_{QCD}\simeq m_q \la\bar{q}q\ra$ in accordance with WI. 

In case of ``deformed QCD" considered in \cite{Thomas:2011ee} we could  explicitly
compute the contact term and see that it is saturated 
by the monopoles which in weak coupling regime describe the tunnelling processes between different topological sectors of the theory. 
While the topological sectors  in case of strongly coupled 4d  QCD of course still exist, we 
  do not have such   luxury to compute the corresponding transitions explicitly. Nevertheless, we can 
  still study the same dynamics of the topological sectors (and transitions between them) by using the ghost as an effective, but not physical, degree of freedom, which leads to expression (\ref{top}). 
  This formula being derived in strongly coupled QCD using the effective ghost description 
  can be compared  with the lattice results, see detail studies of this question in refs.~\cite{Horvath:2005cv,Ilgenfritz:2007xu, Ilgenfritz:2008ia,Bruckmann:2011ve,lattice}.  In particular, one can explicitly see that the singular behaviour of the contact term is not an artifact of any approximation, but an inherent feature of underlying gauge theory. Furthermore, there are  no any physical scale factors (such as $\Lqcd$)  which would determine  the  singular behaviour of this term. In different words,   the  non-dispersive contribution   in  the lattice simulations has diverging nature of the core and vanishing width in the continuum
limit,  in full agreement with our eq. (\ref{top}).
     
     We reproduce Fig.\ref{chi-lattice} from ref.\cite{lattice}  to  illustrate few    elements which are crucial for this work and which are explicitly present on the plot. First of all, there is a  narrow peak around $r\simeq 0$ with a ``wrong sign". Second,  one can observe  a smooth behaviour in  extended region  of  $r\sim \text{few ~fm}$  with the opposite sign. Both these elements  are present
in the lattice computations as one can see from Fig.\ref{chi-lattice}. The same important elements are also present 
in our ghost's based computations given by eq. (\ref{top}).    In different words, the Veneziano ghost does model 
 the crucial  property of the  topological susceptibility related to summation over topological sectors  in gauge theories. This feature  can not be accommodated by any physical asymptotic states as it is related to  non-dispersive contribution  with ``wrong sign"   as explained in the text. Furthermore, as non-dispersive positive contribution has singular behaviour at the core, this behaviour can not be reproduced by any finite size pseuodoparticles  such as finite size instantons in instanton liquid model \cite{Schafer:1996wv}, see more comments on this matter in section \ref{comments}.  
  
  \begin{figure}[t]
\begin{center} 
 \includegraphics[width = 0.6\textwidth]{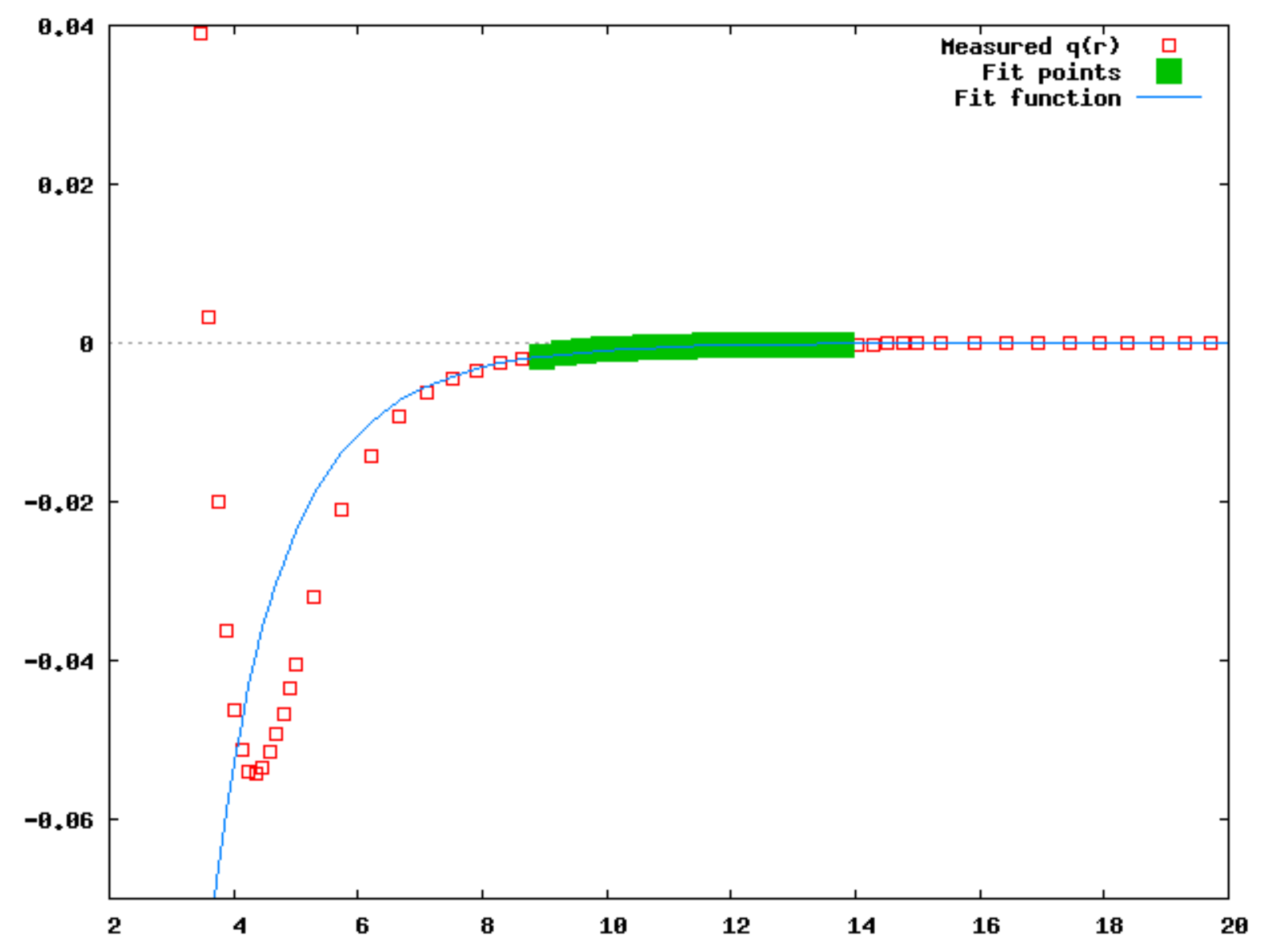}
 \caption{\label{chi-lattice}
 The density of the topological susceptibility $\chi(r)\sim \la q(r), q(0)\ra$ as function of separation $r$ 
such that $\chi\equiv \int d r \chi(r)$, adapted  from~\cite{lattice}. Plot explicitly shows the presence 
of the contact term with the ``wrong sign" (narrow peak around $r\simeq 0$). }
\end{center}
\end{figure}

Most important lesson from these discussions is that the lattice simulations unambiguously show that the low energy relations
discussed in section \ref{theorems}   are not saturated by physical asymptotic states.
Therefore, the effective potential   derived in Section \ref{lagrangian} and representing  these low energy relations also can not be unambiguously 
associated with some physical propagating degrees of freedom (or any their combinations).  
In different words, the axion field $a$ is not   expandable  in terms of some  asymptotic states (pseudoscalar glueballs), 
and can not be described with a canonical kinetic term. 
Though these fields can not be associated with propagating degrees of freedom, the corresponding ``non-dispersive"  fluctuations    are obviously present  in lattice Monte Carlo simulations~\cite{Horvath:2005cv,Ilgenfritz:2007xu, Ilgenfritz:2008ia,Bruckmann:2011ve,lattice}.  

 In what follows we attempt to identify the relevant gauge configurations which make crucial contributions to the low energy relations (\ref{Euclidean-1}) and (\ref{Euclidean-2}). The most important observation from previous section \ref{lagrangian}
 is that the $2\pi$ periodicity of the effective potential hints on possible Euclidean configurations in form of the domain walls which potentially can contribute into the non-dispersive term (\ref{Euclidean-2}). 
In fact,   the lattice results to be reviewed below apparently suggest that such    long range structure indeed emerges, and observed  configurations indeed make crucial contribution  to the low energy relations (\ref{Euclidean-1}) and (\ref{Euclidean-2}). However, the   corresponding configurations can not be  easily described in terms of conventional quantum field theory (QFT) degrees of freedom.
Still, one can argue that the gauge  configurations can be described in terms of the dual  holographic description when the relevant Euclidean configurations can be identified with so-called D2 branes \cite{Gorsky:2007bi,Gorsky:2009me,Verschelde:2011fz,Zhitnitsky:2011aa}. The only element which  is important for us   in what follows is that the tension of these D2 branes vanishes below the QCD phase transition $T<T_c$ such that an arbitrary large number of these objects can be formed\footnote{Vanishing tension in the dual description in the  confined phase is a result of 
 the Hawking-Page phase transition~\cite{wittenterm}  when the  D2 brane shrinks to the tip of a cigar type geometry. It can be interpreted as an instability of a solution.  It can be also interpreted as a formation of tensionless objects. The last interpretation is adapted in the present work. While there are some hints from QFT viewpoint on presence of Euclidean domain wall configurations in the system as we discussed in section \ref{lagrangian}, it is very difficult to describe them  in terms of conventional QFT degrees of freedom.}.  

  \subsection{Insights from numerical  lattice simulations and holographic description: the emergence of coherent structures}\label{coherent} 
 The recent 
Monte Carlo studies of pure glue   gauge theory  have revealed a laminar structure in the vacuum consisting of extended, thin, coherent, locally low-dimensional sheets of topological charge embedded in 4d space, with opposite sign sheets interleaved, see original QCD lattice results~\cite{Horvath:2003yj,Horvath:2005rv,Horvath:2005cv,Alexandru:2005bn}. A similar structure has been also 
observed in QCD by a different group~\cite{Ilgenfritz:2007xu} and also in two dimensional $CP^{N-1}$ model~\cite{Ahmad:2005dr}. Furthermore, the
studies of localization properties of Dirac eigenmodes have also shown evidence for the delocalization of low-lying modes on effectively low-dimensional surfaces. It is not a goal of the present paper to cover this   subject with a number of subtle points which accompany it.  Instead, we limit ourselves by mentioning very few key properties of the gauge configurations which apparently make crucial contributions to the low energy relations (\ref{Euclidean-1}) and (\ref{Euclidean-2}).  Here is the (incomplete) list of the unusual features of these gauge configurations:

a) The tension of the ``low dimensional objects"  vanishes below the critical temperature and these objects percolate through the vacuum, forming a kind of a vacuum condensate;

b) These ``objects"  do not percolate through the whole 4d volume, but rather, lie on low dimensional surfaces $1\leq d < 4$~\cite{Horvath:2005rv};

c) The total area of the surfaces is dominated by a single percolating cluster of ``low dimensional object";

d) The contribution of the percolating objects to the gluon condensate   and therefore to correlation function  (\ref{Euclidean-1})  is opposite in sign compared to its total value;

e) The contribution of the percolating objects to the topological susceptibility (\ref{Euclidean-2}) has the same sign     compared to its total value;

f) The width of the percolating objects apparently vanish in the continuum limit similar to narrow peak of the  lattice size around $r\simeq 0$ plotted on Fig. \ref{chi-lattice}. 

It is very difficult to understand all those properties using conventional quantum field theory analysis. At the same time,
using the  holographic description the interpretation of these Monte Carlo  results fits very nicely with conjecture that the observed   structure 
can be identified with   the D2 branes in holographic description\cite{Gorsky:2007bi,Gorsky:2009me,Verschelde:2011fz,Zhitnitsky:2011aa}. 

In particular the tension of the D2 branes vanishes 
in confined phase  as a result of cigar geometry in the holographic 5d description such that the D2 brane shrinks to the tip as we already mentioned. 
Therefore, an arbitrary large number of D2 branes  could be produced, resulting in their condensation. 
 The observed percolation (condensation) of these objects in lattice simulations unambiguously  imply that they must have vanishing effective tension.
Otherwise, only finite, not infinite percolating  clusters could be observed, in contrast with observations listed as item c). In Minkowski space-time
these Euclidean 4d objects can be  thought as       tunnelling processes  which are taking  place  in vacuum.

Other properties listed above are also have  simple and natural interpretation within holographic framework representing the 
extended, thin, coherent objects  as D2 branes.  In particular,   numerical observation listed as item e) above,  reveals that the contribution of the D2 branes   into the topological susceptibility $\chi_{Eucl}$ given by eq. (\ref{Euclidean-2}) has the same positive sign as the contact term itself.
Furthermore,  both contributions have  opposite sign in comparison with any propagating physical states (\ref{dispersion}). In different words, lattice studies  are consistent  
with    non-dispersive nature  of these objects. 

Another numerical observation listed as item d) above,   reveals that the contribution of the D2 branes  to   the correlation function (\ref{Euclidean-1})  has the positive sign which is the opposite to the total value of the correlation function
 (\ref{Euclidean-1}) which is negative. 
 The presence of this positive contribution indicates that there is a non-dispersive 
 portion in  (\ref{Euclidean-1}). This feature  implies that there is some  ambiguity in standard  procedure 
to saturate    the low energy theorem   
  (\ref{Euclidean-1}) by physical propagating asymptotic states.   

Few comments are in order on interpretation of the D2 branes as Euclidean objects describing the tunnelling processes between degenerate states, see ~\cite{Zhitnitsky:2011aa}
for the detail discussions.   The term ``degeneracy"  
should not  be confused with conventional term ``degeneracy" when two or more physically distinct states are present in the system. In the context of this paper the
     ``degeneracy" implies there existence of winding states $| n\ra$ constructed as follows: ${\cal T} | n\ra= | n+1\ra$.  In this formula the operator ${\cal T}$ is  the  large gauge transformation operator  which commutes  with the Hamiltonian $[{\cal T}, H]=0$ which results in  ``degeneracy" of the winding states $| n\ra$. 
       The presence of $n$ different sectors in the system is reflected  by  summation over $n$ in path integral approach. 
   The property  ${\cal T} | n\ra= | n+1\ra$ is analogous to the Bloch's system in condensed matter physics with the ``only" difference that the states $| n\ra$ are not physically distinct states, and there is no any  real ``physical" degeneracy in the system.  The physical vacuum state in QCD  is {\it unique}
     and constructed as a superposition of $| n\ra$ states.  This interpretation  explains the ``wrong" sign in residues of the correlation function (\ref{Euclidean-2}) in comparison with conventional formula (\ref{dispersion}) as we describe the tunnelling of the Euclidean D2 objects  between winding states $| n\ra$ rather than the tunnelling of conventional physical particles between distinct vacuum states 
     in condensed matter physics in the Bloch's case, see ~\cite{Zhitnitsky:2011aa}
for the detail discussions.

   Such a tunnel-based  interpretation   of D2 branes   is consistent with representation of the contact term (which we assume is saturated by the D2 branes) as a total divergence.  
     Indeed, the non-dispersive contact term with  ``wrong sign" in topological susceptibility   (\ref{top})   can be  represented as a surface integral 
       \bea
   \label{div}
 \chi_{YM} &\sim&  \int   d^4x \left[ \delta^4(x)  \right] =   \int   d^4x~
 \partial_{\mu}\left(\frac{x^{\mu}}{2\pi^2 x^4}\right)= \oint_{S_3}    d\sigma_{\mu}
 \left(\frac{x^{\mu}}{2\pi^2 x^4}\right) \\ \nonumber 
 &=&\frac{1}{24\pi^2}   \oint_{S_3} d\sigma_{\mu}\epsilon^{\mu\nu\lambda\sigma}
 {\rm Tr}[U\partial_{\nu}U^{-1}U\partial_{\lambda}U^{-1}U\partial_{\sigma}U^{-1}], 
    \eea
    where matrix $U$ describes the  topologically non-trivial gauge configurations interpolating between two 
    different winding states $| n\ra$ and $| n+1\ra$.
    Such a representation (\ref{div}) of the  non-dispersive term in topological susceptibility   (\ref{top})  again  supports the tunnelling interpretation  
    of the contact term. The dimensional parameter $ \sim {f_{\eta'}^2 m_{\eta'}^2} $ in eq. (\ref{top}) should be interpreted as a number of tunnelling events per unit time per unit 3d volume in the vacuum. This coefficient    can be computed precisely   in the so-called deformed QCD where all calculations are under complete theoretical control in weakly  coupled  regime,  and where such  interpretation becomes precise   and unambiguous statement ~\cite{Thomas:2011ee}. It is important to emphasize that the singular behaviour of the contact term $\sim \delta^4(x) $ is not an artifact of any approximations, in spite of the fact that this behaviour  was established in simplified models:  in the ``deformed QCD" ~\cite{Thomas:2011ee} and in the Veneziano model (\ref{top}). This singular behaviour and the vanishing width  (in continuum limit) of the contact term    have been observed in the lattice simulations \cite{Horvath:2005cv,Ilgenfritz:2007xu, Ilgenfritz:2008ia,Bruckmann:2011ve,lattice}. 
            
     Also,      
 the lattice measurements  are also consistent with 
   vanishing width of these ``objects", see item f) from the list above.
  These measurements, again,   support the 
tunnelling interpretation of the contact term. Indeed, it is well known property of conventional quantum mechanics that 
the   individual photons penetrate an optical tunnel barrier with an effective group velocity considerably greater than the vacuum speed of light~\cite{Steinberg:1993zz}, see also review \cite{Chiao:1998ah}. In different words, the  
tunnelling  time can not be distinguished  (experimentally)  from zero. In quantum field theory context it means that 
the configurations responsible for tunnelling events could have vanishing size (in continuum), which is precisely what has been measured
on the lattices, see item f) above.  The same feature  can be also seen on Fig. \ref{chi-lattice} where the contact term with vanishing size (in continuum) has a ``wrong sign" in Euclidean lattice simulations. This feature again supports the tunnelling interpretation of this term in Minkowski space.
  \subsection{Few more remarks on long range structure}\label{comments}
  With the structure just described the following question immediately  emerges: what is the relation between ``extended, thin, coherent, locally low-dimensional sheets"  and other well-known topological objects e.g. instantons from the instanton liquid mode (ILM)~\cite{Schafer:1996wv}?  Instantons from ILM are finite size objects, have been  studied on the lattices, and is known to give a considerable contribution to the low energy relations (\ref{Euclidean-1}) and (\ref{Euclidean-2}). The crucial difference is that ILM  has inherent 
  scale, the instanton size such that all correlations decay on this scale $\sim 0.5$ fm. This scale enters all computations, including density-density correlation function shown on Fig.\ref{chi-lattice}.
  The corresponding positive contribution to the peak would have a width of  order the size of the instanton in ILM in contrast with vanishing size of the core observed  in unbiased  lattice  simulations~\cite{Horvath:2005cv,Ilgenfritz:2007xu, Ilgenfritz:2008ia,Bruckmann:2011ve,lattice}. These   lattice studies  were based on new computational technique when  no any auxiliary  procedures such as cooling/smoothing/smearing are used in analysis. At the same time, previously used technique when  cooling/smoothing/smearing   was unavoidable  part of the analysis    is known to be strongly biased towards classical solutions.    In fact, the original papers \cite{Horvath:2003yj,Horvath:2005rv} were precisely devoted to this question with the main conclusion  is  that the topological charge density distribution is not localized in any well -defined finite size lumps, but rather it is strongly delocalized in form of low dimensional objects as discussed above. To conclude: our viewpoint  is that ILM (or any other model based on classical finite size objects) may provide a good estimate for some integral characteristics which are not sensitive to the fine details of topological charge density distribution. Such models would fail in description of some other effects   which are sensitive to local distribution of the topological density and its specific features such as  long range order.
  
    Our next comment is as follows. The discussions presented above suggest that the relevant structure can not be described in terms of   semiclassical configurations  with finite size and tension (such as instantons/sphalerons etc). Can one visualize the relevant objects
    discussed above in section \ref{coherent}, at least qualitatively? In fact, an analogous  domain wall structure is known to exist in QFT at large temperature in weak coupling regime where it can be described in terms of classical equation of motion.  These are  so-called   $Z_N$ domain walls 
    which separate domains characterized  by a different value for the Polyakov loop at high temperature. As is known, see e.g. recent review 
    paper \cite{Fukushima:2011jc} and references therein, these  $Z_N$ domain walls interpolate between topologically different but physically identical states    connected by large gauge transformations similar to our discussions at the end of section \ref{lagrangian}. At high temperature these objects can be described in terms of classical equation of motion. In this regime they  have finite tension $\sim T^3$ such that their contribution to path integral is strongly suppressed. While the corresponding topological sectors are still present in the system at low temperature (though they are realized in a different way) it is not known how to describe the fate of $Z_N$  walls within QFT     in strong coupling regime when semiclassical approximation breaks down. From holographic perspective however, the corresponding domain walls can be identified with  tensionless D2 branes, and the hope is that such objects can be understood within holographic description  as sketched  in section
    \ref{coherent}.

   \section{Applications: From Heavy Ion Collisions to Cosmology}{\label{applications}
   \subsection{Basic idea}\label{idea}
    In previous sections \ref{lagrangian} and   \ref{lattice}  we presented a number of arguments 
   suggesting that the low energy relations (\ref{Euclidean-1}), (\ref{Euclidean-2}), and therefore the low energy potential (\ref{potential}), are saturated  by  coherent,  extended, thin gauge configurations. The arguments were based on recent lattice results \cite{Horvath:2003yj,Horvath:2005rv,Horvath:2005cv,Alexandru:2005bn,Ilgenfritz:2007xu} and some model considerations.   We interpreted these configurations as  tunnelling processes which are happening all the time in Minkowski vacuum with no  interruptions~\cite{Zhitnitsky:2011aa}. These tunnelling processes in Minkowski vacuum (when no any external sources are present in the system) do not lead to any emission or absorption of real particles, similar to the persistent tunnelling events in Bloch's case.  These tunnelling events simply
   select an appropriate ground state of the system which is a specific superposition of $| n\ra$ states. In what follows we assume that these coherent,    gauge fluctuations   saturate the low energy relations, and we  concentrate only on  these configurations. As we shall see below, the long range structure of the vacuum fluctuations plays a crucial role in both applications considered below. If this structure were not present in the system, both effects  would be negligibly small. In different words, all conventional models with finite correlation length would predict a negligibly small magnitude for the effects considered below. The long range order  advocated in this work is the key element fundamentally different from   all other models  characterized by finite correlation length  as discussed in section \ref{comments}.  
   
   The effective potential  (\ref{potential}) which gives us a hint that such long range order is present in the system 
   (as a result of $2\pi$ periodicity of the potential supporting the  long range domain walls) in this set up should be interpreted as follows.  
   While each gauge configuration  has definite sign of the topological charge density, the  opposite sign sheets alternate.  This {\it delicate cancellation} between  the  opposite sign sheets leads to the known result corresponding to   minimization of  the effective potential in physical vacuum:   $\la a \ra_{min}=0, ~\la \eta\ra_{min}=0$. This is exactly the result obtained by conventional integrating out  $G^2$ and $G\tilde{G}$ fields as described in section~\ref{lagrangian} at $\theta=0$. In different words, the total vacuum expectation value  of $\cal{P}$-odd  operator $\la a \ra\sim \la G\tilde{G} \ra =0$ vanishes as it should at $\theta=0$.  
   
   However, when    some external sources  are present in the system  a {\it delicate cancellation} between  the  opposite sign sheets may lead,   in general,  to a local minimum with nontrivial 
   values for the dilaton and the axion fields  in the region where external impact is felt, i.e. $\la a(x) \ra\neq 0, ~\la \eta(x) \ra\neq 0$.
     Non-vanishing and long range correlated   $\la a(x) \ra\neq 0$ obviously implies that $\la G\tilde{G}\ra \neq 0$ in the same region of space-time. 
  Indeed,  the axion field $\la a \ra\neq 0$  plays the role of effective $\theta_{ind}\neq 0$ as  can be seen from definition (\ref{pot}).
 Precisely this deviation from   vacuum values (\ref{min}), as we shall argue below, will be the main source for  the   local violation of $\cal{P}$ invariance in QCD as well as for   the ``apparent universal thermalization"  in high energy collisions. The long range order advocated in this work 
 plays a crucial role is estimation of a magnitude of the effects. All conventional models based on instantons/sphalerons  with finite correlation length would lead to negligible  effects.

   We consider   two applications   of these ideas,  which are separated by gigantic  differences in scales. First, in section \ref{collisions} we consider high energy collisions when Minkowski vacuum is disturbed with a typical $\Lambda_{QCD}\sim 10^2$ MeV scale.   Secondly, in section \ref{cosmology}  we consider cosmology
   when the deviation from Minkowski vacuum is characterized by  the Hubble constant  $H\sim 10^{-33}$ eV. These two 
   applications with drastically different scales have a unique  crucial element  in common:  in both cases the Minkowski vacuum is deviated from its equilibrium dynamics, and the observable effects are proportional to this deviation. This disturbance, of course,  is numerically very different: in first case the external influence leads to the effects of order of one, while in the second case the effect is $\sim H/\Lambda_{QCD}\sim 10^{-41}$. Still, in both cases, all observational  effects are due to the same ``non- cancellation" between   long range  vacuum    configurations  with opposite sign sheets interleaved. 
    This ``non-cancellation"  in both cases is a direct consequence of   an external    disturbance  of the  vacuum  from its  perfect   state $\la a \ra=0, ~\la \eta\ra=0$ in infinitely large Minkowski  space time. Normally,  in unperturbed vacuum, these  long range  vacuum    configurations are subject of delicate and exact cancellation between   low-dimensional opposite sign sheets  of topological charge. This exact cancellation  leads to $\la G\tilde{G}\ra = 0$ in the vacuum state.     Our key assumption is that    this exact cancellation can not been maintained in new   environments considered below in sections~\ref{collisions} and \ref{cosmology}. Furthermore, this ``non-cancellation" occurs in extended region of space-time with the correlation length being the same order of magnitude as size of vacuum sheet itself, which  is exactly the source of long range effects discussed  below.

  \subsection{Terrestrial applications: high energy collisions at RHIC and LHC.}\label{collisions}
 In the framework presented  in sections \ref{lagrangian} and   \ref{lattice}   we argued that the  
 long range correlated,  thin gauge configurations with vanishing width make a crucial  contribution 
to   the low energy relations (\ref{Euclidean-1}), (\ref{Euclidean-2}), and therefore to the low energy potential (\ref{potential}).
In Minkowski space time the Euclidean objects with vanishing sizes should be interpreted as the tunnelling events 
which are happening with superluminal velocity, see last paragraph in section \ref{lattice}. 
\exclude{  In fact, the superluminal behaviour has been experimentally observed in a number of tunnel experiments, see e.g.    ~\cite{Steinberg:1993zz} when it has been found  that  an effective group velocity considerably greater than the vacuum speed of light for individual photons penetrating an optical tunnel barrier, see also review \cite{Chiao:1998ah}.  We note that causality is not violated in these tunnelling events as any information/signal can not be transferred.  
In quantum field theory context as argued above
the configurations responsible for tunnelling events must have vanishing size (in continuum), which is precisely what has been measured on the lattices. }
  Now we present some arguments suggesting that   precisely these type of configurations with vanishing width (in Euclidean space) and interpreted as  instantaneous superluminal tunnelling events (in Minkowski space) are responsible for the origin of local violation of $\cal{P}$ invariance as well as for the ``apparently" thermal spectrum in high energy collisions. 

The basic picture behind  this proposal is that high energy collisions  including $e^+e^-, ~pp$ and $p\bar{p}$  interactions do not completely destroy the coherent vacuum structure described  
in section \ref{lattice}. Instead, the collisions lead to some shifts and distortions between low-dimensional sheets  (which existed before the collision)   rather than   breaking them. These distortions, however,   spoil the  exact cancellations between  the  long range  vacuum    configurations which existed in vacuum before the collision occurred.  
It is crucial that these Euclidean long range  vacuum    configurations with vanishing width actually describe instantaneous tunnelling events in Minkowski space. 
 Distortion of these configurations 
 leads to emission of real physical particles. It is similar to the  inelastic scattering processes of quasi-electrons in the presence of  impurities   distorting   the perfect lattice structure  in Bloch's case   in condensed matter physics.  
In this proposal  two crucial puzzles formulated in section \ref{T} are immediately resolved as follows:

{\bf a)}  The thermal spectrum in $e^+e^-, ~pp$ and $p\bar{p}$ high energy collisions emerges in spite  of the fact that the statistical thermalization could  never be reached in those systems. This is due to the fact that this spectrum is a result of the same tunnelling processes which have been happening all the time even before the collision occurs.  However,   in the vacuum these tunnelling processes (between ``degenerate"  $|n\ra$ states)  
do not lead to any emissions as a result of the {\it delicate cancellations} which is a consequence of exact symmetry 
$[{\cal T}, H]=0$. 
The collision slightly disturbs  the vacuum and leads to formation of a locally  different 
$|\theta_{ind}\ra =\sum \exp(in\theta_{ind})|n\ra$ states with quasi-momentum $\theta_{ind}$ constructed from the same  ``degenerate"  and correlated $|n\ra$ states. These $|\theta_{ind}\ra $ states decay as a result of tunnelling processes leading to the thermal spectrum of emitted  particles. The spectrum could be nothing else but thermal as it is entirely due to the tunnelling events with superluminal velocity when information can not be transferred\footnote{It is interesting to note that the computations  based on the  Euclidean configurations, e.g. instantons,   describing the tunnelling events,  also produce a Planck like spectrum\cite{Ostrovsky:2002cg, Shuryak:2002an,Basar:2012jb}. We interpret this result as a generic feature  of physics of tunnelling  rather than a simple  coincidence.}.   In holographic picture, the observed spectrum is a result of emissions  from fluctuating  D2 branes which are disturbed by collisions. 

We must emphasize that we concentrate  on low energy part of the spectra in this paper; the high part comes from many  different other processes which shall not be even mentioned here. Furthermore, in heavy ion collisions a conventional statistical thermalization is achieved, in contrast with ``apparent thermalization" discussed in this work. This part of the spectra, being characterized by a different   so-called saturation $Q_{sat}$ scale, is also not a part of consideration of the present work.

{\bf b)}  An approximate  universality of the temperature   with no dependence on energy of colliding particles nor their nature 
(including $e^+e^-, ~pp$ and $p\bar{p}$ collisions) is due to the fact that the emission occurs from the distorted   QCD vacuum state represented by   the  long range  vacuum    configurations with vanishing width rather than from the colliding particles  themselves.  In holographic picture  the observed spectrum is a result of emissions 
from the disturbed  D2 branes; therefore,   it can not be sensitive to a nature of disturbance and   always remains the same. The intensity of the emission, of course, depends on the nature of colliding particles, and total energy being transferred to the D2 branes to excite them from their normal equilibrium state in Minkowski space-time. 

\exclude{ as the collisions do not completely destroy the coherent vacuum structure  which eventually responsible for the thermal spectrum.}

Along with this simple resolution of   two aforementioned puzzles   a) and b),  this proposal shed some light on  another puzzle formulated in section \ref{P}: why the $\cal{P}$ odd domains are so large, much larger than conventional $\Lambda_{QCD}^{-1}$ scale? Resolution of this puzzle as given   in item c) below also elucidates  the reason  why we treat two naively different problems formulated in sections \ref{P} and \ref{TT}  as two sides of the same coin.

{\bf c)}  A puzzle with a  long- range  structure of the  $\cal{P}$ odd domains within our framework  is resolved as follows. The low-dimensional coherent sheets responsible for the tunnelling as explained above, carry 
 the quantum numbers of the topological charge density $G\tilde{G}$ which is $\cal{P}$ and $\cal{CP}$ odd operator. 
 In fact, it was exactly    this feature which was  studied   on the lattices \cite{Horvath:2003yj,Horvath:2005rv,Horvath:2005cv,Alexandru:2005bn,Ilgenfritz:2007xu}.
  Therefore, the distortion of these low-dimensional coherent sheets due to the collisions lead to  a local ``non-cancellation" 
 between different low-dimensional coherent sheets. Precisely this structure becomes  coherent on the large scales, much larger than 
$\Lambda_{QCD}^{-1}$ as a result of collisions. In different words, the collisions do not produce a coherent large $\cal{P}$ odd   domain. Rather, the collisions  do not completely destroy the coherent structure which always existed in vacuum. 
The role of collisions in this framework is that the collisions  slightly destroy the  {\it delicate cancellation} which is inherent   feature of the  perfect undisturbed vacuum state as was discussed  in section \ref{idea}.

 This picture gives a precise dynamical realization of  the conjecture 
formulated in \cite{Zhitnitsky:2010zx} that these two phenomena outlined    in sections \ref{P} and \ref{TT} are in fact are intimately related as both originated from  the dynamics of the coherent vacuum structure observed on the lattices  and described  
in section \ref{lattice}. The crucial point here is this: though we can not presently compute the spectrum, it  must demonstrate the same features   for $\cal{P}$ even as well as for  $\cal{P}$ odd correlations with universal  Planck spectrum observed in all high energy collisions. Some   supporting evidence for this behaviour is  listed in  items f), g) h) below.  In holographic description both these phenomena formulated in sections \ref{P} and \ref{TT}  are due to the same tunnelling events described by D2 branes which emit real particles as a result of small disturbance of the vacuum state resulted from the collision.

      With this basic picture outlined above the main question is: 
 What are the technical tools   to describe these effects quantitatively? 
 As we argued in section \ref{comments} the relevant structure can not be described in terms of semiclassical physics 
 with finite correlation length  using ILM or sphalerons as relevant pseudoparticles. Instead, one should 
   study the dynamics using the dual  holographic description. To be more specific,  one should analyze  the dynamics  of tensionless D2 branes when they are slightly disturbed by external forces (including the fluctuations of D2 branes leading to the  emission of real particles). 
In lattice simulations, the relevant information is hidden in dynamics of extended coherent vacuum sheets when they are slightly disturbed by the collisions. The fluctuations of the corresponding correlated coherent objects are expected to emit particles with  thermal spectrum.  In principle, these ideas can be tested using the lattice simulations. These are  technically very challenging problems, which are beyond the scope of the present work. 

 Still, there are some  model independent consequences of this framework which are based exclusively  on the assumption that the collisions  do not completely destroy the coherent structure, but rather slightly disturb it. These generic consequences will be listed below.  We leave the subject with  more specific but  model dependent calculations for a future  work. 
Here  we   continue the   list   of model-independent consequences of this framework:

{\bf d)}  When a system is not infinite, but sufficiently large (e.g. large  ions with size $L\gg \Lqcd^{-1}$) the observable  $\cal{P}$ odd  effect due to the collisions of objects size $L$ are expected to be  proportional to $1/L^p$ with some power $p$.  The  $L^{-p}$ scaling essentially describes the deviation of the system from  the  ground state in the region  $L$ as a result of collision.   
We refer to  Appendix \ref{L} where we present a number of arguments  (including some     QCD lattice results) 
 supporting $L^{-p}$  scaling.  Such a Casimir like behaviour should be contrasted with naively expected exponential suppression $\exp(-L)$ when a mass gap being inherent feature of QCD is present in the system. The crucial point is that  $L^{-p}$ correction 
 is originated from non-dispersive contributions  which are   not  related to any    physical states  as discussed in section \ref{lattice}. 
  
 {\bf e)} As one can see from Fig. \ref{RHIC}   some suppression of the measured correlations  with increasing  the size of the system indeed has been observed. We would like to  interpret this suppression as a manifestation of the $L^{-p}$    scaling. 
   Indeed, the effect for ${\text{Au+Au }} $ collisions  with $A\simeq 197$ is obviously suppressed in comparison with 
 ${\text{Cu+Cu }} $ collisions with   $A\simeq 64$. There are   many other effects which influence this ratio (both systems are obviously not very large when derivative expansion is justified). However, the effect goes in the right direction (the effect is stronger for 
  a smaller size system ${\text{Cu+Cu }} $  than for a  larger  ${\text{Au+Au }} $ system).  
    
 \begin{figure}[t]
\begin{center}
 \includegraphics[width = 0.4\textwidth]{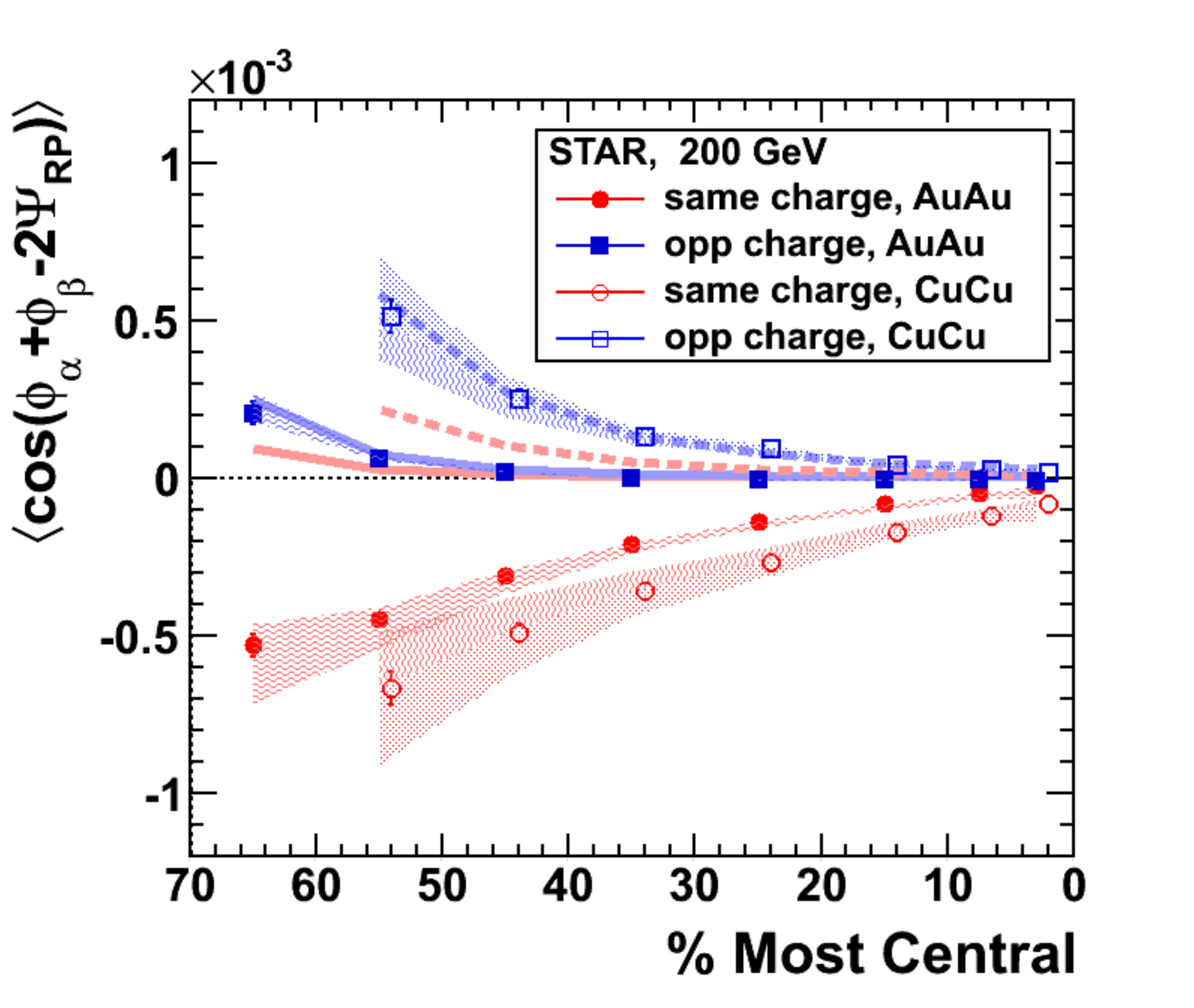}
  \includegraphics[width = 0.4\textwidth]{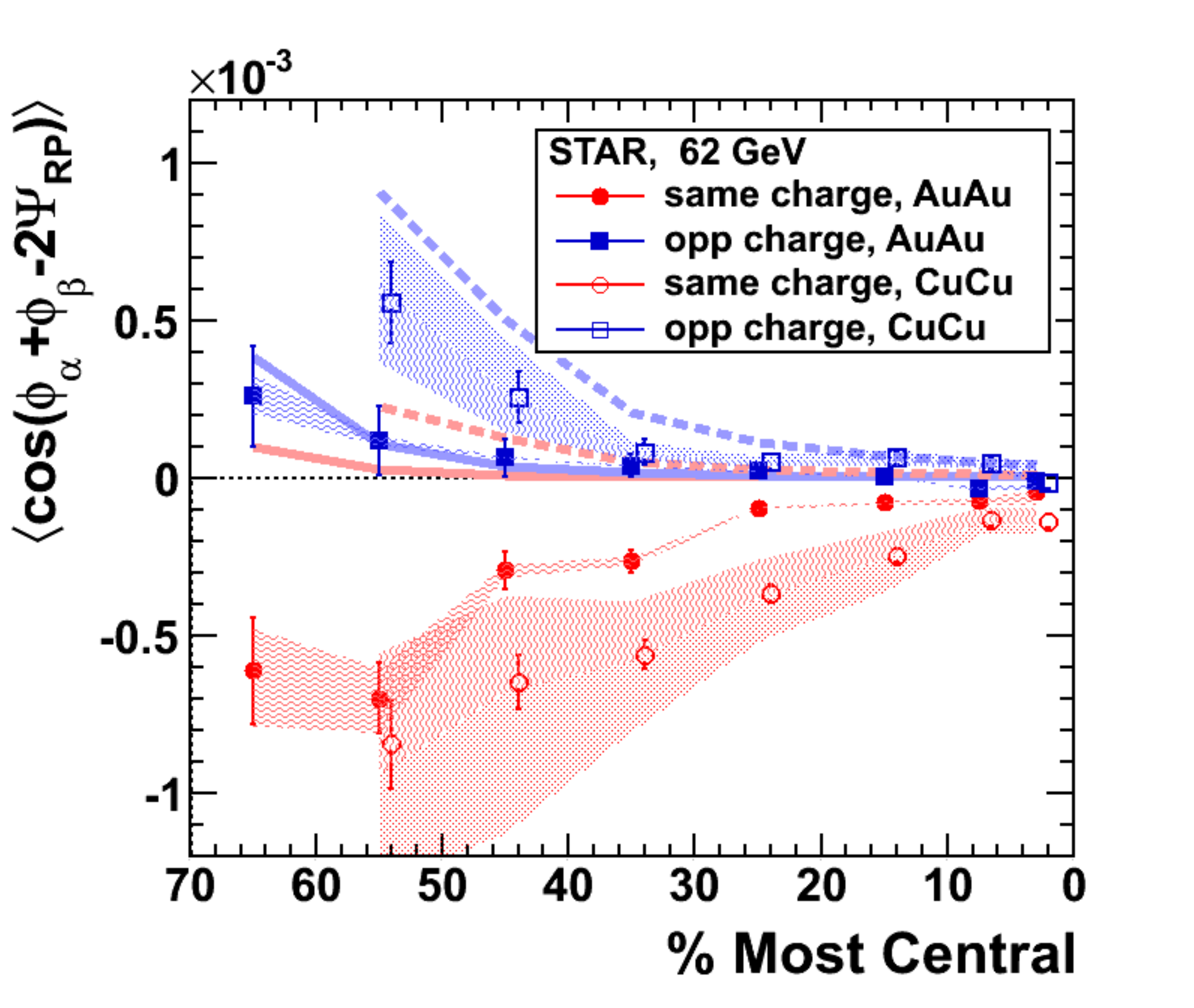}
 \caption{\label{RHIC}
Data for $\sqrt{s_{NN}}=200\text{GeV}$ and $\sqrt{s_{NN}}=62\text{GeV}$  for Au+Au and Cu+Cu collisions  (adapted  from~\cite{Abelev:2009tx}). 
The plots demonstrate the universality in behaviour, see items d) and e) in the text for the details.}
\end{center}
\end{figure}

       {\bf f)}  The correlations due to the local $\cal{P}$ violation  should demonstrate the universal behaviour similar to the ``universal apparent  thermalization"  discussed in section \ref{TT} as the source for the both effects  is the same as argued in this paper. In particular, the effect should not depend 
       on energy of colliding ions.
        Such independence on energy is indeed supported by observations        where correlations measured in ${\text{Au+Au }} $ and ${\text{Cu+Cu }} $ collisions at $\sqrt{s_{NN}}= 62~ {\text{GeV}} $ and  $\sqrt{s_{NN}}= 200~ {\text{GeV}} $  
       are almost identical and  independent on energy, see Fig. \ref{RHIC}. 
       
               \begin{figure}[t]
\begin{center}
  \includegraphics[width = 0.4\textwidth]{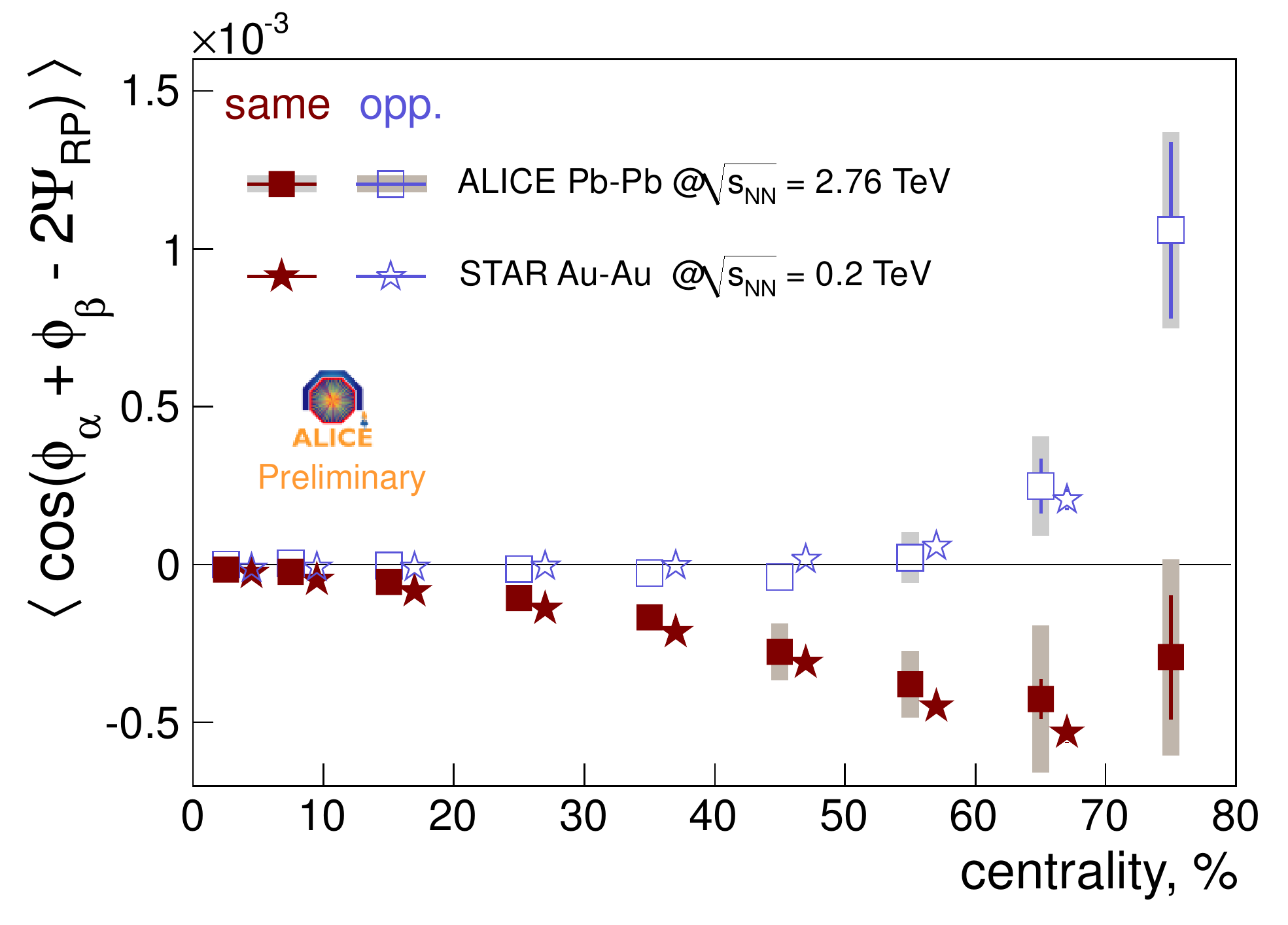}
 \caption{\label{LHC}
Data for Pb-Pb collisions with ALICE at the LHC  at $\sqrt{s_{NN}}=2.76~\text{TeV}$ and Au+Au collisions with STAR at RHIC at $\sqrt{s_{NN}}=0.2~\text{TeV}$  (adapted  from~\cite{Selyuzhenkov:2011xq}). 
The plot demonstrates the universality   in behaviour (energy independence, $2.76~\text{TeV}$ vs $0.2~\text{TeV}$), see item g) in the text for the details.}
\end{center}
\end{figure}

        \begin{figure}[t]
\begin{center}
 \includegraphics[width = 0.4\textwidth]{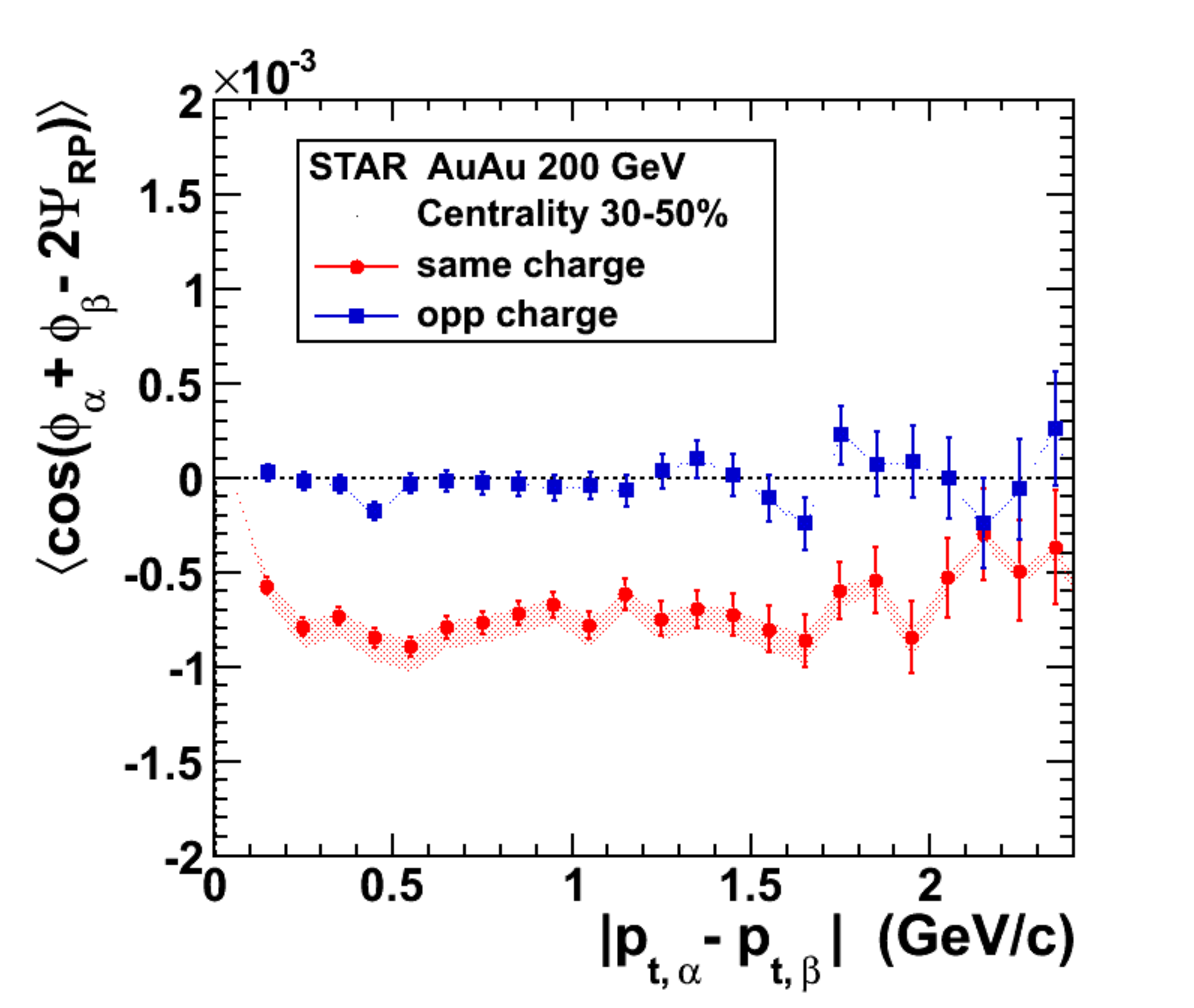}
  \includegraphics[width = 0.4\textwidth]{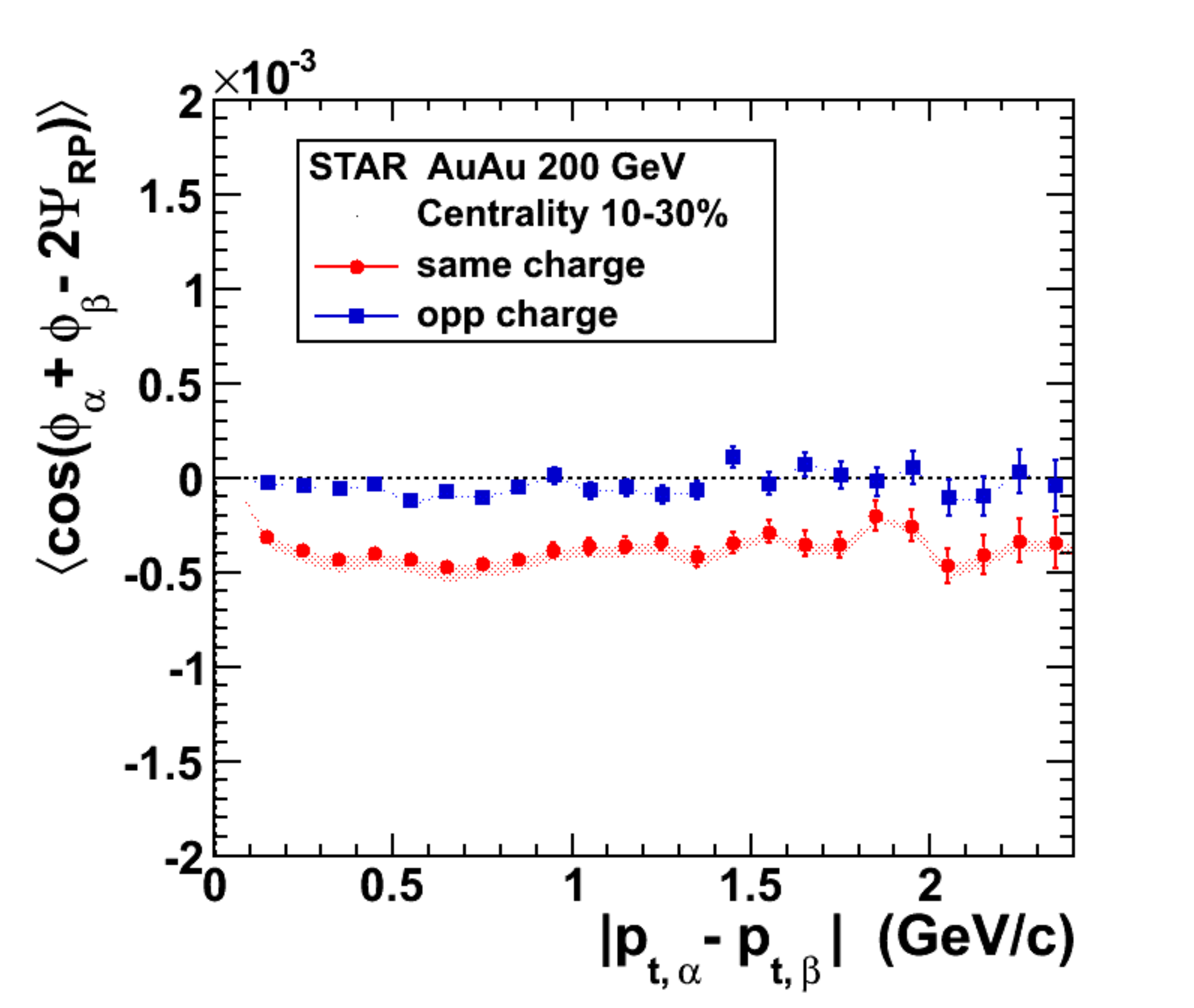}
 \caption{\label{k}
Data for $\sqrt{s_{NN}}=200\text{GeV}$ and $\sqrt{s_{NN}}=62\text{GeV}$  for Au+Au and Cu+Cu collisions  (adapted  from~\cite{Abelev:2009tx}). 
The plots demonstrate the universality in behaviour and independence on $|k_{\perp, \alpha}-k_{\perp,\beta}| $, see item h) in the text for the details.}
\end{center}
\end{figure}

 {\bf g)} We expect the same tendency to continue for the LHC energies. In fact, 
in \cite{Zhitnitsky:2010zx}  we had predicted (before the LHC preliminary results ~\cite{collaboration:2011sma,Selyuzhenkov:2011xq} have been posted) that the corresponding correlations at the LHC energies should  demonstrate a similar strength and a similar
       features found at RHIC. Preliminary recent results from ALICE Collaboration indeed 
       confirm this prediction. As one can see from Fig \ref{LHC}, 
 the  ALICE ${\text {Pb-Pb}}$ results for 3-particle  correlator at $\sqrt{s_{NN}}= 2.76~ {\text{TeV}} $ are 
       almost identically coincide with RHIC   ${\text{Au+Au }} $ results  at $\sqrt{s_{NN}}= 200~ {\text{GeV}} $,
       which is precisely what was anticipated as the sizes for ${\text{Au}^{79}} $ and ${\text{Pb}^{82}} $ are almost the same
       in contrast with much lighter and smaller ${\text{Cu}^{29}} $ when effect  should be stronger, see item e) above.

   {\bf h)}  The arguments presented above on universality of the   correlation strength 
     do not depend  on transverse momenta $k_{\perp}^2$. This   is a consequence  of  the   same universal behaviour  discussed above.  Indeed, the  low-dimensional coherent sheets responsible for the tunnelling/emission   are the ${\cal P}$ odd objects as they  carry the quantum numbers of the topological charge density.  Therefore, the corresponding emissions   always contribute to the correlations presented   on Fig. \ref{k}   with the same rate as for conventional Hagedorn emission. This consequence  of the universality  is also  consistent with observations, see Fig~\ref{k}, where it is found that the correlation depends very weakly on $|k_{\perp, \alpha}-k_{\perp,\beta}| $.
     The signal (intensity of the asymmetry) shown in Fig. \ref{k}  obviously becomes weaker for central collisions, as $\cal{P}$ odd correlations are washed out for the central collisions, see discussions below on relevant  time scales for this problem. However, the dependence on transverse momenta $k_{\perp}^2$ is determined by the same tunnelling processes leading to the Planck spectrum, and therefore must be universal. We should emphasize once again that we are talking about the low energy part of the spectra  where our consideration is applicable,  see the corresponding comment in item a) above.  \\

       {\bf Few more comments.} The qualitative consequences which follow from the picture outlined above are  consistent with all presently available data. Even more that that, few predictions we had made in \cite{Zhitnitsky:2010zx}  
        have been recently confirmed at the LHC energies, see item g). The basic idea   is  that all effects  are   
        proportional to the deviations  of the axion $ a (x) $ and $  \eta (x) $ fields from their vacuum values computed in infinitely large space-time (\ref{min}).      Small disturbances  resulting from collision      can be parametrized  by   $|\la\partial_{\mu}a(x)\ra |\sim |\la\partial_{\mu}\eta(x)\ra |\sim L^{-1}$ which is a measure of deviation from unperturbed ground state in the large region of size $L$.   
        The deviations should be  computed from their vacuum values: $\la a \ra = \theta/N_c, ~ \la \eta\ra=0$          presented in eq.(\ref{min}). In a simplest case of slow time- dependent and spatially-  independent variation of the axion and dilaton fields $a(t),~ \eta(t)$, the key parameter is the acceleration $| \la\dot{a}\ra | \sim |\la\dot{\eta}\ra |\sim  |\mathbf{a}| $, see Appendix \ref{a}. Derivative expansion which is assumed in this work can be justified only for small acceleration 
       $ \mathbf{a}\ll \Lqcd$.

     In reality  $\mathbf{a}$ is not numerically  small number as estimations  of ref. \cite{Castorina:2007eb} suggest.
   Furthermore,   the  size of  ${\cal P}$ odd domain can  not be very large even for non-central collisions for physically available ions. The finite size effects  and other non-universal features may lead to some corrections from the universal picture presented here. The crucial element for   analysis of  these non-universal features   is the understanding of all relevant time scales of the problem.
   \exclude{ which, in principle, determined  by the dynamics of the $a(x), \eta(x)$ fields 
   governed by potential (\ref{potential}).} In particular, if for a given acceleration $\mathbf{a}$   a typical time scale  $\Delta t$ for multiple  emissions  satisfies the condition  $\Delta t \gg \mathbf{a}^{-1}$, there will be no any observable $\cal{P}$ odd effects as the system has time to  completely  adjust  to a new environment caused by a collision\footnote{It is similar to more familiar case of fluctuating instantons (sphalerons) and antiinstantons which are $\cal{P}$ odd objects themselves. However, these  $\cal{P}$-odd  fluctuations of the topological charge do not lead to any observable $\cal{P}$ odd effects because of the cancellations between small instantons and antiinstantons. Difference with our case is that the relevant objects in our analysis are strongly  correlated, thin extended gauge configurations discussed in section \ref{lattice}. It is precisely these long range correlations may lead to ``non- cancellation" between two types of  low-dimensional sheets observed on the lattices, and identified with D2 branes in holographic description.}. To observe  $\cal{P}$ odd effects the opposite condition $\Delta t \leq \mathbf{a}^{-1}$ must be satisfied. This hierarchy 
     of scales also explains the dependence of strength of correlations from centrality as presented in Fig.\ref{RHIC}.  
     Indeed, the peripheral collisions obviously correspond to a smaller disturbance of the ground state and correspondingly smaller acceleration  $\mathbf{a}$, while $\Delta t \sim \Lambda_{QCD}^{-1}$ is essentially determined by conventional QCD scale and does not much depend on centrality. This is precisely the reason why the condition $\Delta t \leq \mathbf{a}^{-1}$
     is likely to  be met in heavy ion peripheral collisions. Strength of the correlation obviously increases with centrality 
    where acceleration $\mathbf{a}$ decreases,  in accordance  with data presented  in Fig.\ref{RHIC}. 
  
      The  condition $\Delta t \leq \mathbf{a}^{-1}$ is less likely to be satisfied    for centralities  $(0\%-5\%)$ in heavy ion collisions as well as  for       $e^+e^-, ~pp$ and $p\bar{p}$ collisions, see  \cite{Zhitnitsky:2010zx} with more discussions on this point. In this case, one should expect a   conventional Planck emission   from disturbed low-dimensional extended coherent topological sheets as usual. However, the asymmetry presented in Figs. \ref{RHIC}, \ref{LHC} in this case will be largely washed out because too many disturbed sheets contribute to the emission with opposite signs. 
      
    Also, when total energy becomes sufficiently small, a typical momentum transfer $\Delta q$  of emitting particles is also getting smaller. It leads to increasing of  typical $\Delta t \sim (\Delta q)^{-1}$ such that condition $\Delta t \leq \mathbf{a}^{-1}$ is no longer satisfied, and effect is washed out again. This time it is washed out   not because of sufficiently large $\mathbf{a}$, typical for centralities $(0\%-5\%)$, but rather because of relatively large $\Delta t \sim (\Delta q)^{-1}$ as a result of relatively small momentum transfer $\Delta q$.\\

 {\bf Order of magnitude estimate.} 
We are now in position to give a simple, order of magnitude, estimation for all  $ {\cal P}$ odd effects when all deviations from the vacuum are small, 
$| \la\dot{a}\ra | \sim  |\mathbf{a}| $, see Appendix \ref{a}.
  The simplest way to proceed is   to  use the description in terms of the Veneziano ghost which effectively describes the dynamics of the degenerate topological sector of QCD.   The 
         number density  of the ${\cal P}$   odd domains with size $\lambda\simeq \frac{2\pi}{\omega}$ can be estimated as follows, see \cite{Zhitnitsky:2010zx} for the details
           \bea
\label{N_ghost}
 d N_{\omega} =   \frac{d^3k}{(2\pi)^3} \frac{2 }{(e^{2\pi\omega/\mathbf{a}}-1)}.
\eea
  The total contribution to the energy associated with these soft fluctuations is  
      \bea
\label{E_ghost}
 E_{ghost}\simeq \int \frac{d^3k}{(2\pi)^3} \frac{2\omega }{(e^{2\pi\omega/\mathbf{a}}-1)}=\frac{\pi^2}{15}\left(\frac{\mathbf{a}}{2\pi}\right)^4, 
\eea
which  should be compared with conventional    contribution due to  $N_f$ massless  quarks  and $N_c^2-1$ gluons, 
     \bea
\label{gluons}
 E_{q+g} \simeq \frac{\pi^2}{15}  \left(\frac{\mathbf{a}}{2\pi}\right)^4\left[ (N_c^2-1) +\frac{  7N_cN_f}{4}\right].
\eea
      Therefore, the relative energy associated with slow ghost fluctuations with $0^{+-}$ quantum numbers in comparison with conventional   fluctuations of quarks and gluons with Hagedorn spectrum 
      is estimated to be
      \bea
      \label{kappa_1}
      \kappa\equiv \frac{ E_{ghost}}{ E_{q+g}}\sim \frac{1}{\left[ (N_c^2-1) +\frac{  7N_cN_f}{4}\right]},
      \eea
      which is numerically $\sim  0.05 $.  The effect is parametrically small at large $N_c$ and  proportional $\sim 1/N_c^2$ which is  a typical suppression  for any phenomena  related to topological fluctuations. The effects related to the disturbance of the well organized structure 
      described in section \ref{lattice},  obviously vanish at $\mathbf{a}=0$ as eq. (\ref{E_ghost}) states. Non-vanishing $\mathbf{a}\neq 0$
 effectively describes the dynamics of  excited ``topological sectors" as discussed in \ref{idea}. 
      The factor $  \kappa$ essentially counts number of fluctuating degrees of freedom which lead to the ${\cal P}$ and ${\cal CP}$ odd  environment. However, these degrees of freedom are not the asymptotic states, and they  do not propagate
      to infinity, and they do not contribute to the absorptive parts of any correlation functions.  Rather, they contribute to the non-dispersive parts of the correlation functions as explained in section \ref{qft}. 
      
    There are few more   factors which must be present in estimate for the observed    asymmetries  presented on Figs. \ref{RHIC}, \ref{LHC}, \ref{k}. First, there is a trivial numerical factor  proportional to the electric charge $e\sim \sqrt{\alpha}\sim 10^{-1}$. Second, 
    there is an additional suppression factor $\sim L^{-1}$   as  a manifestation of the Casimir scaling discussed in the text and   in Appendix \ref{L}. This is the key element of the entire framework: all observed asymmetries would be much smaller if the long range order is not present in the system. We parametrize the Casimir behaviour $\sim L^{-1}$ as follows
     \bea
      \label{final}
      \text{[asymmetries  on Figs. \ref{RHIC}, \ref{LHC}, \ref{k}]} \sim e\cdot \kappa \cdot \frac{f(\gamma)}{L\Lqcd}&\sim& 5\cdot 10^{-4}, \\ \gamma \equiv \mathbf{a}\Delta t, ~~ f(\gamma\ll 1)\simeq const, ~~  f(\gamma\gg 1)&\simeq& 0, \nonumber
      \eea
where for numerical estimates we assume $L\Lqcd \simeq 10$. Numerical estimation (\ref{final}) is consistent with intensities of the observed asymmetries presented  on Figs. \ref{RHIC}, \ref{LHC}, \ref{k}. The dimensionless parameter $\gamma\equiv \mathbf{a}\Delta t$ is a convenient way to parametrize  different   time-scales  discussed above. 
 Function $f(\gamma)$ in expression (\ref{final})   vanishes at large $\gamma\gg 1$ and approaches a constant at small $\gamma\ll 1$ as argued above. The dependence on centrality is effectively represented by variation of  parameter  $\mathbf{a}$, and therefore $\gamma$.  Function $f(\gamma)$ in expression (\ref{final}) also depends on many other characteristics of the system such as charge/size of a nucleus, its induced magnetic field,  and many other non-universal  parameters. Evaluation of function $f(\gamma)$ is a hard problem which is beyond the scope of the present work. The corresponding computations  would obviously require  some specific   model- dependent assumptions, while we attempt in this work to formulate some model-independent consequences of this framework (see items {\bf a-h} above) which are  based exclusively on the basic fundamental principles formulated in the beginning of this section, and not on any additional model-dependent/non-controllable approximations. 
 
  If a long range order would not be present in the system (for example, we would use a finite size sphaleron  transition to estimate the effect), we would 
  get a strong suppression $\exp(-\Lqcd L)$ instead of power like suppression ${(L\Lqcd)}^{-1}$ from eq. (\ref{final}). 
  Such a strong $\exp(-\Lqcd L)$ suppression is a consequence of a finite correlation length ${(\Lqcd)}^{-1}$ typical for all conventional models.
 It  should be contrasted 
  with framework advocated in this work when the long range order is inherent feature of the system.
  Therefore, any estimates based on conventional finite size objects would unavoidably   lead to an estimate which is   $ \exp(-\Lqcd L)\cdot (L\Lqcd)\sim (10^{-3}-10^{-4})$    smaller than the observed values. 
       
       We should emphasize once again that  all measurements   are  ${\cal P}$ even observables. Therefore,  there are many background processes  which considerably contribute to the effect
    \cite{Bzdak:2009fc,Liao:2010nv,Bzdak:2010fd,Schlichting:2010qia,Pratt:2010zn}.  If future studies of this effect nevertheless confirm   that  the CME/charge separation effect  is a main  source of the observed event-by-event fluctuations, we would like to interpret this  result as a strong evidence for the long range order present in the system, as advocated in this proposal.    Much work needs  to be done before a qualitative picture sketched  above  becomes a quantitative description of  the ${\cal P}$ and ${\cal CP}$ odd
     correlations observed at RHIC and LHC~\cite{Voloshin:2004vk,Abelev:2009tx,collaboration:2011sma,Selyuzhenkov:2011xq}.

  \subsection{Cosmological applications: dark energy and accelerated universe}\label{cosmology}

In this section we consider another application related to the deviation of the ground state from its 
constant  values (\ref{min}) as a result of expansion of our FLRW universe. This deviation is  determined by the Hubble constant  $H\simeq 10^{-33} {\text {eV}}$ which replaces acceleration parameter $\mathbf{a}\simeq 10^2~{\text {MeV}}$   describing the deviation of the QCD ground state   resulting from a  collision, see previous section~\ref{collisions}. This study 
may have some profound consequences on our understanding of FLRW universe we live in.   What is more remarkable is the fact that the fundamental cosmological ideas formulated below, in principle, can be experimentally tested in heavy ion collisions at RHIC and  LHC where the required unusual environment  can be produced.
  
Non-dispersive contribution with a ``wrong sign" in topological susceptibility (\ref{top}) obviously implies,  as eq. (\ref{5})  states, that there is also some energy 
related to this contact term.
This  $\theta-$ dependent portion of the  energy not related to any physical propagating degrees of freedom,  is well established phenomenon and tested on the lattice, see Fig.\ref{chi-lattice}; it is not part of the debates.
What is the part of the debates and speculations 
is the question on how this energy changes when  background varies. In different words, the question is: how does   the non-dispersive contribution to the $\theta-$ dependent portion of the  energy vary when conventional Minkowski background is replaced by FLRW universe with the horizon size $L\sim H^{-1}$ determined by the Hubble constant $H$?
Similar problem  has been already discussed in section \ref{collisions}, where we argued that the entire effect 
of  the collisions can be described as  the deviation from the ground state values (\ref{min}) and must be proportional 
to $L^{-p}$ when disturbance is  relatively small   for sufficiently large $L$, see item d) in previous section.
 
The motivation for this question in the present context is different as there are no any external sources which may change the properties of the ground state, similar to collisions in section \ref{collisions}. Instead, the deviation from solution (\ref{min}) emerges as our  
universe is not static, but rather, expanding with the rate $H$.   
We adapt the paradigm that the relevant definition of  the energy   which enters the Einstein equations 
 is the difference $\Delta E\equiv (E -E_{\mathrm{Mink}})$, similar to the well known Casimir effect when the observed   energy is in fact a difference 
  between the energy computed for a system with conducting boundaries  and infinite Minkowski space.     This   is  in fact  the standard subtraction procedure  which is normally used for description the horizon's thermodynamics \cite{Hawking:1995fd,Belgiorno:1996yn} as well as  in a course of computations of different Green's function in a curved background by subtracting infinities originated from the flat   space~\cite{Birrell:1982ix}. 
  In the present context  such a definition $\Delta E\equiv (E -E_{\mathrm{Mink}})$ for the vacuum energy for the first time was advocated   in 1967   by Zeldovich~\cite{Zeldovich:1967gd} who argued that  $\rho_{\text{vac}} \sim Gm_p^6 $ with $m_p$ being the proton's mass, see also
  \cite{Zhitnitsky:2011tr} for related references. In different words, the dark energy observed  in our universe might be  a result of mismatch between the QCD vacuum energy computed in slowly expanding universe with the expansion rate $H$  and the one which is computed in flat Minkowski space.

This is exactly the motivation for question formulated in the previous paragraph:  how does $\Delta E$ scale with $H $? 
The difference $\Delta E$  must obviously vanish when $H\rightarrow 0$ as it corresponds to the transition to flat Minkowski space. How does it vanish? A naive expectation based on common sense suggests that 
$\Delta E \sim \exp(-\Lambda_{QCD}/H)\sim \exp(-10^{41})$ as QCD has a mass- gap $\sim \Lambda_{QCD}$, and therefore, $\Delta E$ must not be sensitive to size of our universe\footnote{Here and in what follows we   emphasize on the power like sensitivity to arbitrary large distances irrespectively to  their nature. In different words, a crucial distinct feature which characterizes the system  we are interested in is the presence of dimensional parameter $L\sim H^{-1}$   in a system which discriminates it  from infinitely large  Minkowski space-time. For purposes of this work we do not discriminate   the horizon size $H^{-1}$ of expanding universe 
from   size $L\sim 10~$fm of a compact 
manifold we used in  previous section \ref{collisions}.}. Such a naive expectation formally follows from the dispersion relations similar to (\ref{dispersion})
which dictate that a sensitivity to very large distances must be exponentially suppressed when  a mass gap is present in the system. 

However, as we discussed at length in this paper, along with conventional dispersive contribution  (\ref{dispersion}) there is also the non-dispersive contribution (\ref{Euclidean-2}, \ref{top}) which plays a crucial role in the construction of the ground state (\ref{min}). 
This term,  as explained in   section~\ref{lattice} is not related to any propagating physical degrees of freedom, and 
  may lead to a power like 
scaling $\Delta E\sim H +{\cal O} (H)^2$ rather than naively expected  $\Delta E \sim \exp(-\Lambda_{QCD}/H)$ behaviour.

The Casimir  power like scaling $\Delta E\sim H +{\cal O} (H)^2$  in QCD, if confirmed by future analytical and numerical studies, may have profound consequences 
  for understanding of the expanding FLRW universe we live in.     If true, the difference between two metrics (FRLW and Minkowski) would lead to an estimate 
   \beq
   \label{Delta}
   \Delta E\sim H\Lambda_{QCD}^3\sim (10^{-3} {\text eV})^4,
   \eeq
which is amazingly close to the observed DE value today without adjusting of any parameters. 
Such a behaviour $\Delta E\sim H +{\cal O} (H)^2$   was   postulated in \cite{UZ}.   
 The power like  scaling has received recently a solid theoretical and numerical support, see    Appendix \ref{L}  for references and the details.
      
     It is interesting to note that 
expression (\ref{Delta})  reduces to 
Zeldovich's \cite{Zeldovich:1967gd} formula   $\rho_{\text{vac}} \sim Gm_p^6 $ 
   if one replaces $ \Lambda_{QCD} \rightarrow m_p $   and $  H\rightarrow G \Lambda_{QCD}^3$. 
   The last step follows from 
      the  solution of the Friedman equation 
     \beq
 \label{friedman}
 H^2=\frac{8\pi G}{3}\left(\rho_{DE}+\rho_M\right),  ~~ \rho_{DE}\sim H\Lambda_{QCD}^3, ~~ \rho_c=\frac{3H^2}{8\pi G}
  \eeq    
  when the DE component dominates the matter component, $ \rho_{DE}\gg\rho_M$. In this case    the evolution of the universe  approaches a  de-Sitter state with constant expansion rate $H\sim G \Lambda_{QCD}^3$ as follows from (\ref{friedman}).
  
   One should add that a number of other fine tuning issues which always plague dark energy models, such as ``coincidence problems'', ``drastic separation of scales'', ``unnatural weakness of interactions'', etc.,  possess a simple and natural explanation within this framework without a single new field/coupling constant in the fundamental Lagrangian of Standard Model.
In particular, the fine tuning problem which goes under the name of ``cosmic coincidence''  problem finds its natural resolution as follows.
The   vacuum energy which attributed to the extra Casimir -type contribution (\ref{Delta}) in our framework   becomes relevant when its  energy is of the same order of magnitude as the critical density,  $\rho_{DE} \simeq \rho_c $.  Equating these two quantities returns $t_0 \sim H^{-1}\sim (G \Lambda_{QCD}^3)^{-1}\sim 10~ $Gyr, see eq. (\ref{friedman}). This is indeed a  correct estimate for  the lifetime of  the present universe.

   To conclude this section: the key element for both applications considered in sections \ref{collisions} and \ref{cosmology} is the paradigm that the observed effects are due to the disturbances of the ground state  (\ref{min}).  The sources for these perturbations,  of course, are very different: in first case it is a result of heavy ion collision, while in the second  case it is a result of expansion of the universe. However, in both cases  this deviation 
   from uniform solution (\ref{min}) demonstrates   a Casimir power like correction to the vacuum energy in spite of the fact that QCD is a confined theory with a gap. Formally, the Casimir  type behaviour emerges from  non-dispersive contact term which is
   not related to any propagating degrees of freedom, but rather, originated from degenerate topological sectors of the theory. Therefore,  a standard  argument based on dispersion relations suggesting the  exponentially weak sensitivity to arbitrary large distances is simply not applicable in this case.   Microscopically this long-range order emerges  as a result of dynamics of   low-dimensional coherent sheets of  gauge configurations  seen in the lattice simulations and   discussed  in section \ref{lattice}.  We refer to Appendix \ref{L}
   with more insights and discussions on this  very nontrivial Casimir-like behaviour. 
  
   \section{Conclusion}
   
   The main results of this work can be formulated in few lines as follows. We formulated the paradigm that two naively unrelated phenomena: local $\cal{P}$ violation  observed in heavy ion collisions   and universal thermal aspects observed in all high energy collisions are in fact both related to quantum anomalies of QCD.  The basic idea is that   
   the well established low energy relations   representing the quantum anomalies are saturated by extended, thin, coherent, locally low-dimensional sheets of topological charge embedded in 4d space, with opposite sign sheets interleaved.  High energy collisions 
   disturb this well-organized structure and lead to emission of physical particles. We argued that a number of long standing puzzles are immediately resolved within this framework. In particular, the thermal spectrum in   high energy collisions emerges in spite  of the fact that the statistical thermalization could  never be reached in those systems.
   Also: an approximate  universality of the temperature   with no dependence on energy of colliding particles nor their nature 
 is due to the fact that the emission occurs from the distorted   QCD vacuum state  rather than from the colliding particles  themselves.  Finally, as the low-dimensional coherent sheets  carry 
 the quantum numbers of the topological charge density, they are responsible for a long range order of the $\cal{P}$  odd domains, see section \ref{collisions} for the details and comparison with available data. 
 
 The key point in the analysis of the $\cal{P}$  odd correlations is that the distortion of the ground state follows the Casimir  type scaling $\sim L^{-p}$ for peripheral collisions for sufficiently large  size of the system. This scaling should be contrasted with naive expectation which  predicts the    exponential type behaviour $\sim \exp(-\Lambda_{QCD}L)$. Precisely this Casimir  type scaling, in our view, leads to observed, sufficiently strong $\cal{P}$ odd correlations (\ref{final}), while for  exponential type behaviour
 all correlations would be washed out, and numerically much smaller than observed. 
 
 The main reason why the Casimir  type scaling emerges in a gapped theory such as QCD is the presence of a  non-dispersive contribution which is
   not related to any propagating degrees of freedom, but rather, originated from degenerate topological sectors of the theory. The power like behaviour $\sim L^{-p}$ is, of course, another manifestation of the same long range low-dimensional coherent sheets which are observed on the lattices, and  which saturate the low energy relations. 
   
   We interpret a relatively large observed intensity of the correlations (\ref{final}) as a direct manifestation of the Casimir-like scaling and long range order in the QCD vacuum, as 
   these   correlations should be strongly suppressed if one takes into account only conventional sources characterized by a finite correlation length $\sim (\Lqcd)^{-1}$.
   Future experiments at RHIC and LHC should confirm or rule out this proposal based on the idea  of  the  low-dimensional coherent sheets 
   which are always present in the QCD vacuum as lattice simulations suggest.

  It  is quite  remarkable   that some of   fundamental properties of QCD, such as  Casimir  type scaling and long range order, may have profound consequences for cosmology when a relatively small (in QCD units) parameter $ L^{-1} \sim (10~ {\text{fm}})^{-1}$ from application to collisions (\ref{final})    is replaced by drastically smaller Hubble constant $H\sim 10^{-33} \text{eV}$ in FLRW universe from cosmological application leading to estimate (\ref{Delta}).   The DE  in this framework emerges as mismatch  between the energies  of a system in a non-trivial FLRW background  and Minkowski space-time geometry,  similar to the well known Casimir effect.  In this framework the   DE is entirely rooted into   the  strongly coupled QCD, without any new fields and/or coupling constants.  What is even more remarkable is that some of the most intricate properties of the quantum ground state  can be, in principle, experimentally tested in heavy ion collisions at RHIC and LHC, where such unusual environment can be achieved. This idea can be also tested using the lattice Monte Carlo simulations which would allow to study the  DE on a computer. 
     
  \section{Acknowledgements}
       
       This study had started    during  the workshop ``the first heavy ion collisions at the LHC"  at CERN, August 2010,   where work \cite{Zhitnitsky:2010zx} had been presented. I had received a numerous number of questions during and after my talk. The basic question was on a relation between formal computations in terms of the acceleration $\mathbf{a}$ and description in terms of the gauge fields of underlying gauge theory. This work is the first step to shed some lights on this relation.
        I am    thankful to   a number of people including  Edward Shuryak, Berndt Mueller,  Andrey Leonidov and other participants of the   workshop   for initiating this study. I am also  thankful to    James Bjorken for many hours of discussions during his visit to  Vancouver.    I am also thankful to Ivan Horvath and Folk Bruckmann for correspondence on lattice results and comments  on feasibility to measure the Casimir like correction in Monte Carlo simulations. 
 I am also    thankful to  Dima Kharzeev, Larry McLerran,   and other members of Nuclear Physics groups at BNL and  Stony Brook U.     for useful and stimulating discussions related to the subject of the present work.
 This research was supported in part by the Natural Sciences and Engineering Research Council of Canada.

\appendix
\section{  On relation of  the axion field $a(x)$ with other relevant QCD parameters}\label{a}

We wish to understand   qualitative behaviour of the system 
  when the ground  state    only slightly deviates from its undisturbed  structure   as a  result of collision.
   Formally, it corresponds   to a study of the system when    $  a(t, \vec{x}), \eta(t, \vec{x})$ vary  very slowly, i.e. we use the derivative expansion. 
   We make an additional simplifying assumption that the system is sufficiently large such that   $  a(t, \vec{x}), \eta(t, \vec{x}) $ do not actually depend on
   spacial coordinates $\vec{x}$. In this case the problem is reduced to an accounting for  the first non-vanishing terms in derivative expansion proportional to $  \dot{a}$ and $  \dot{\eta}$ which  we take as phenomenological  parameters for our problem.
For this simplified case  we want to argue that slow varying axion and dilaton fields $a(x), \eta(x)$ are related to the acceleration $``\mathbf{a}"\neq 0$  
and some other  parameters describing the external sources, see below. 

Indeed, any local disturbances 
resulted from the collision should be described in terms of derivatives $\partial_{\mu}a(x)$ rather than field $a$ itself. This is because $\la a\ra = \rm{ const.}$ corresponds to a solution for a minimum of the potential (\ref{min}) with  non-vanishing $\theta$ in entire space, while we are interested in local variations due to the collisions. To simplify things further we consider only time dependent variation $\sim \la\dot{a}\ra$ assuming that there is no any  coordinate dependence $\vec{x}$ in the system, which corresponds to the collisions of infinitely large ions. For slowly varying axion field in adiabatic approximation  one can identify  $\la\dot{a}\ra={\dot{\theta}_{ind}}/{N_c}$
as it represents a minimum for non vanishing  $\theta = \rm{ const.}$ as computed in ref.\cite{Halperin:1997bs,Halperin:1998rc}.

Furthermore, $\dot{\theta}_{ind}=\mu_L-\mu_R$ can be identified with the difference of chemical potentials of the right
$\mu_R$ and left $\mu_L$ handed fermions if they are present in the system, see  \cite{Zhitnitsky:2010zx,Kharzeev:2009fn} 
for the details on this identification. In holographic description this parameter is expressed in terms of the boundary condition
$\dot{\theta}_{ind}=A_0(z=\infty)$ for the Ramond-Ramond gauge  field $A_0(z)$ defined in the bulk, see \cite{Brits:2010pw} for the details on this identification.

Finally,   one can argue that parameter  $\dot{\theta}_{ind}$ should be proportional to the acceleration $\mathbf{a}$
when the theory is formulated in geometrical terms. 
The argument goes like this. We adapt the conjecture formulated in \cite{Castorina:2007eb}   that apparent thermalization 
observed in high energy collision experiments is a result of tunnelling.    In our framework, the quantum tunnelling  is  happening all the time in vacuum 
(even before the collision occurs) between the topological sectors, while the observed emission  is a result of external impact  which 
causes  a  non-cancellations between topological extended  sheets    responsible for the tunnelling,  as described in section \ref{idea}.
     Nevertheless, if  these tunnelling events can be reformulated eventually in geometrical terms as advocated in \cite{Castorina:2007eb}, the only small parameter which characterizes a small  deviation of the system from Minkowski space time is the acceleration $\mathbf{a}$. In conventional QFT description the only parameter which describes the deviation of the system from Minkowski  vacuum 
      (\ref{min})  is $\dot{\theta}_{ind}$. When both parameters are small, and they both describe the same physics, and they both describe the deviation from perfect   Minkowski space time  solution (\ref{min}), they must be proportional to each other, i.e. $\mathbf{a}\sim \dot{\theta}_{ind}$.     Another argument leading to the same conclusion  was presented in  \cite{Zhitnitsky:2010zx}  and 
was based on description of the $\theta_{ind}$ state in terms of the  Veneziano ghost's fluctuations in accelerating frame
which saturate the non-dispersive term in low energy relations as discussed in section \ref{qft}. 
The basic idea  is that a typical wavelength $\lambda$ of the $\cal{P}$ odd fluctuation due to the  ghost's fluctuations in the accelerating frame is of order $\lambda\sim 2\pi/\mathbf{a}$ expressed in terms of  acceleration $\mathbf{a}$. A similar scale but  expressed in terms of time variating external parameter 
$ \dot{\theta}_{ind}\sim \omega/(2\pi)=\lambda^{-1}$ leads to a desired relation $\dot{\theta}_{ind}\sim \mathbf{a}/(2\pi)$.

 Therefore, in slow varying environment we have the following parameters which all describe the deviation from undisturbed  vacuum state (\ref{min}) and proportional to each other, 
\beq
\label{identifications}
\la\dot{\eta}(x)\ra\sim \la\dot{a}(x)\ra \sim \frac{\dot{\theta}(x)_{ind}}{N_c}\sim \frac{\mathbf{a}(x)}{2\pi N_c}\sim \frac{[\mu_L(x)-\mu_R(x)]}{N_c}
\sim \frac{A_0(z=\infty)}{N_c} \sim constant.
\eeq
Precisely these parameters should play a crucial role in description of the disturbed vacuum state as a result of collision. 
These parameters in principle accumulate entire  information about low energy relations, contact terms, non-dispersive contributions, including the information about the tunnelling events represented by summation over 
$n$-sectors in path integral and expressed in terms of a ``wrong sign" in topological susceptibility (\ref{Euclidean-2})
in comparison with conventional contribution (\ref{dispersion}). In case of slowly variational background all these parameters
can be approximated by a constant (\ref{identifications}).  Therefore, we can use all our previous results formulated in terms of the acceleration \cite{Zhitnitsky:2010zx}  if  it is sufficiently  small, $| \la\dot{a}\ra | \sim |\la\dot{\eta}\ra |\sim  |\mathbf{a}|   \ll \Lambda_{QCD}$. 
In this approximation one can  explicitly see emergence of the  thermal spectrum as a result of the Unruh radiation~\cite{Birrell:1982ix} when  acceleration $\mathbf{a}$ is treated as a constant parameter. We refer to   paper \cite{Zhitnitsky:2010zx} 
where all relevant formulae in the present context with emerging 
  Planck spectrum have been discussed. As we mentioned in the text, the Planck spectrum is a generic feature when the emission is a result of the tunnelling when no information can be transferred. It is very satisfying to see how this spectrum is emergent in the adiabatic  limit  when the axion field, $\dot{a} \sim \mathbf{a}/(2\pi N_c)$ varies slowly.
  
  Proportionality of all observables    to the acceleration $| \la\dot{a}\ra | \sim |\la\dot{\eta}\ra |\simeq  |\mathbf{a}| \sim T$    is essentially  the same effect as Casimir like suppression $L^{-1}$ discussed in the text and Appendix \ref{L}.
  Both parameters describe  small  deviations along the spatial coordinates $L_x\sim L_y\sim L_z\sim L$ or along 4-th coordinate $T^{-1}$, and both parameters $T$ as well as   $L^{-1}$ are assumed to be small $\ll \Lqcd$ in derivative expansion.  In relativistic system these parameters are essentially the same. 
 In principle, the dynamics  of these phenomenological parameters is governed by the dynamics of D2 branes as we discussed in section \ref{lattice}. The D2 branes  start to fluctuate as a result of the collision, and a small ``non-cancellation" effects are effectively described by a small parameter $\mathbf{a}$.

\section{On Casimir like behaviour  $1/L$ in QCD}\label{L} 
There are a number of arguments supporting the Casimir type    behaviour $\Delta E\sim L^{-p} $ in QCD demonstrating a huge sensitivity to arbitrary large distances in spite of the presence of a gap in QCD. The difference $\Delta E\equiv (E -E_{\mathrm{Mink}})$ here is defined as the difference in energies computed for the system with nontrivial geometry/boundaries  and  infinite Minkowski space, similar to the   well known example of the Casimir effect.  
This behaviour should be contrasted with conventional computations of the Casimir effect for a massive scalar particle with mass $m$  which  leads to the  expected exponential scaling 
 $\Delta E\sim\exp (-mL)$, see e.g.\cite{Casimir}. 
 However, as we discussed in the text, along with conventional dispersive contribution in the system, there is also the non-dispersive contribution (\ref{top})   which emerges as a result  of topologically nontrivial sectors in four dimensional QCD. This contact term may lead to a power like 
corrections $ L^{-p} $ rather than exponential like $  \exp(-\Lqcd L)$ because  the dispersion relations do not dictate the scaling properties of this term. In different words, a power like 
behaviour  $ L^{-p} $ may emerge 
 despite the presence of a mass gap in the system.
  Such a power-like correction  obviously  is very  unexpected effect which   begs for a simple intuitive explanation.  
   
  First example 
   is an explicit computation~\cite{Urban:2009wb} in exactly solvable  two- dimensional QED   defined in a box size $L$. The model has all elements crucial for present work: non-dispersive 
  contact term   which emerges  due to the topological sectors of the theory.
   This model    is known to be  a  theory of a single physical massive field. Still, 
one can explicitly compute  $\Delta E \sim L^{-1}  $ which is in drastic contrast with  naively expected exponential suppression, $\Delta E\sim e^{-L}$~\cite{Urban:2009wb}.  
One more support   in  power like behaviour is an  explicit computation in a simple case of  Rindler space-time in four dimensional QCD  
~\cite{Zhitnitsky:2010ji, Zhitnitsky:2010zx, ohta} where Casimir like correction have been computed using  unphysical Veneziano ghost which effectively describes the dynamics of the topological sectors and the contact term when the background is slightly modified. Thus, power-like behaviour is not a specific feature of two dimensional physics as some people (wrongly) interpret the results ~\cite{Urban:2009wb}.

Power like behaviour  $\Delta E \sim L^{-p}$  is also supported by  recent lattice results \cite{Holdom:2010ak}.
    The approach advocated in ref.\cite{Holdom:2010ak}  is based on physical Coulomb gauge  when nontrivial topological structure of the gauge fields is represented by the so-called Gribov copies leading to strong infrared singularity. In different words, the same Casimir- like scaling 
    emerges in a different framework where  unphysical Veneziano ghost (used in refs. \cite{Zhitnitsky:2010ji, Zhitnitsky:2010zx, ohta})   is  not even mentioned.  
    
    Similar conclusion also follows from``deformed QCD" model computations where $\Delta E \sim L^{-1}  $ correction emerges  in a theory where weak coupling regime is enforced by construction and all computations are under complete theoretical control~ \cite{Thomas:2012ib}.
The very same conclusion also follows from the holographic description of the contact term presented in \cite{Zhitnitsky:2011aa}. The key element for this conclusion follows from the fact that the contact term 
    in holographic description is determined by massless Ramond-Ramond (RR) gauge field defined in the bulk of 5-dimensional space.
 Therefore, it is quite  natural to expect that massless R-R field in holographic description leads to   power like corrections  when the background is slight modified.

     As this  effect plays a crucial role in the applications considered in section \ref{applications},  we want to present  here few other systems    where a similar phenomena occurs, and  where it has precisely the same nature.  Furthermore, in these systems a similar  problem  can be exactly solved (in drastic contrast with  strongly coupled 4d QCD). Most importantly, an analogous   effect  in these systems has been   experimentally observed. 
  
 We start  from   the well known  Aharonov -Casher effect as formulated in \cite{Reznik:1989tm}.
The relevant part of this work can be stated as follows. If one inserts an external charge into superconductor when the electric field is exponentially suppressed $\sim \exp(-r/\lambda)$ with $\lambda $ being the penetration depth,  a neutral magnetic fluxon will be still sensitive to an inserted external charge at arbitrary large distance. The effect is pure topological and non-local in nature.  The crucial element why this phenomenon occurs in spite of the fact that the system is gapped is   the presence of different topological states    in the system.  
 We do not have a luxury to solve a similar  problem   in  strongly coupled four dimensional QCD analytically. However,  one can argue that  the role of ``modular operator"  of   \cite{Reznik:1989tm}  
 (which is the key element in demonstration of long range order)  is played by large gauge transformation operator $\cal{T}$ in QCD which also commutes with the hamiltonian $[{\cal T},H]=0$, such that our system must be  transparent to   topologically nontrivial pure gauge configurations, similar to transparency of the superconductor to the ``modular electric field", see \cite{Zhitnitsky:2011aa} for the details. 
Such a behaviour of our  system can be thought as a non-local topological effect
  similar to the   non-local Aharonov -Casher effect. The last word whether this analogy can be extended to the strongly coupled four dimensional QCD remains, of course,   the prerogative of the direct lattice computations.
We should mention that there are few other systems, such as topological insulators, where a topological long range order emerges in spite of the presence of a gap in the system.  We refer to ref \cite{Zhitnitsky:2011aa} for relevant  references and details.

\end{document}